\documentclass{aastex63}
\usepackage{graphicx,subfigure,color,amsmath,pifont,multirow,threeparttable}
\usepackage{colortbl}
\usepackage{bm}
\definecolor{mycyan}{rgb}{1.00,0.91,0.86}

\submitjournal{ApJ}
\shorttitle{Emission polarization of the internal-shock} 
\shortauthors{Zhang et al.} 

\newcommand{\MyFigA}{\ref{MyFigA}}
\newcommand{\MyFigB}{\ref{MyFigB}}

\newcommand{\MyFigD}{\ref{MyFigD}}
\newcommand{\MyFigE}{\ref{MyFigE}}
\newcommand{\MyFigF}{\ref{MyFigF}}
\newcommand{\MyFigG}{\ref{MyFigG}}
\newcommand{\MyFigH}{\ref{MyFigH}}

\begin{document}
\title{Revisit the Emission Polarization of the Internal-shock for the blazars' Jet}
\correspondingauthor{Da-Bin Lin}     
\email{lindabin@gxu.edu.cn}   
\author{Hao-Qiang Zhang}
\affil{Guangxi Key Laboratory for Relativistic Astrophysics, School of Physical Science and Technology, Guangxi University, \\Nanning 530004, China}
\author{Da-Bin Lin}
\affil{Guangxi Key Laboratory for Relativistic Astrophysics, School of Physical Science and Technology, Guangxi University, \\Nanning 530004, China}
\author{Kuan Liu}
\affil{Guangxi Key Laboratory for Relativistic Astrophysics, School of Physical Science and Technology, Guangxi University, \\Nanning 530004, China}
\author{En-Wei Liang}
\affil{Guangxi Key Laboratory for Relativistic Astrophysics, School of Physical Science and Technology, Guangxi University, \\Nanning 530004, China}


\begin{abstract}
Recent Imaging X-ray Polarimetry Explorer (IXPE) observations of blazars tend to support the shock model for the X-ray emission, but report a low polarization degree ($\Pi\sim 10\%$) in X-rays
compared with the previous theoretical expectations in the shock model.
In order to reconcile the theoretical expectations with observations,
we revisit the polarization of the shock emission by
considering different kind of direction distribution for the shock-generated magnetic fields (sgMFs).
Here, $w'_{\rm sg}\propto(\sin\theta')^{\zeta_{\rm sg}}$ with $\theta'=0$ along the shock normal direction is used to describe the direction distribution of the sgMFs in the shock co-moving frame.
It is found that the polarization in the X-ray and radio emission
for a general jet in blazars
can be described as
$\Pi\sim 44.5[1-\exp(-\zeta_{\rm sg}/2.6)]\%$ and $\Pi\sim 20[1-\exp(-\zeta_{\rm sg}/2.4)]\%$, respectively.
Correspondingly, one can have $\zeta_{\rm sg}\sim 1-1.5$ according to the IXPE observations.
Besides the sgMFs,
the magnetic fields generated by the Richmyer-Meshkov instability (rmMFs)
is supposed to present in the jets.
The direction of the rmMFs is mainly distributed along the shock normal in the simulations
and thus $w'_{\rm rm}\propto(\cos \theta')^{\zeta_{\rm rm}}$
is adopted to describe the direction distribution of rmMFs.
We find that the rmMFs is likely to significantly affect the polarization properties at the low-frequency emission, especially when the sgMFs decay rapidly.
Based on the contemporaneous radio and X-ray observations,
we find the the emission of the electrons in the rmMFs make a significant contribution in the low-frequency emission
and  the the ordered background magnetic fields (obMFs) can be neglected.
\end{abstract}

\section{Introduction}\label{Introduction}
Blazars are the radio-loud subclass of active galactic nuclei (AGN; \citealp{1978bllo.conf..328B}; \citealp{1979ApJ...232...34B};\citealp{2019ARA&A..57..467B}; see \citealp{2019ARA&A..57..467B} for a review),
which power a relativistic jets point at a small angle $\theta_{\rm obs}$ to the Earth's line of sight \citep{1995PASP..107..803U, 1996A&AS..120C.481S} and exhibit a broad two-hump structure in their characteristic spectral energy distributions (SEDs; \citealp{1999APh....11..159U}).
The low-energy hump (from radio to X-ray frequency range) is generally considered as the synchrotron emission of the relativistic electrons in the jet (\citealp{1998MNRAS.301..451G}), while the high-energy hump (from X-ray to $\gamma$-ray bands) is usually believed to originate from synchrotron self-Compton or hadronic process (\citealp{2001ApJ...554..725T, 2011ApJ...736..131A, 2013ApJ...768...54B}).
Blazars usually show the extreme variability in their light curves \citep{2019NewAR..8701541H}, accompanied by the significantly/obviously linearly polarization in multi-waveband observations \citep{2018MNRAS.473.1850A, 2021MNRAS.501.3715B, 2022Natur.611..677L, 2022ApJ...938L...7D, 2023ApJ...942L..10M, 2023ApJ...948L..25P, 2023MNRAS.523.4504O}.


Multi-band polarimetric measurements have been considered as a effective tool to break the degeneracies in the SEDs modelling between the leptonic model (\citealp{1992ApJ...397L...5M, 1994ApJ...421..153S}) and hadronic model (\citealp{2013ApJ...768...54B,2015MNRAS.448..910C}), and to explore the magnetic field configures of the emission region \citep{1988ApJ...332..678J, 2014ApJ...780...87M, Tavecchio_F-2018-Landoni_M-MNRAS.480.2872T, 2020MNRAS.498..599T, 2021Galax...9...37T}.
In previous works, it is believed that the variations in the flux and linear polarization of blazars were largely stochastic in nature and this phenomenon could be explained by the turbulence magnetic field in the emission region \citep{2014ApJ...780...87M, 2021Galax...9...27M}.
However, some recent works (\citealp{2017MNRAS.472.3589K, 2018MNRAS.474.1296B}) have found that the radio and optical polarization properties, such as the polarization angle (PA) rotations, are not simply consistent with a purely random process.
It is indicated that the magnetic field is not purely stochastic in the emission region,
and parts of magnetic field could be in order (\citealp{2018ApJ...864..140P, 2008Natur.452..966M, 2021Galax...9...27M}).
The ordered magnetic field component could exist in the shock compressed large-scale background magnetic field \citep{2005MNRAS.360..869L, 2012AJ....144..105H, 2016ApJ...817...63Z}, the shock generated turbulence (\citealp{Tavecchio_F-2018-Landoni_M-MNRAS.480.2872T, 2020MNRAS.498..599T, 2021Galax...9...58G}), or in a kink-instability-induced magnetic reconnection environment \citep{2021MNRAS.501.2836B, 2021ApJ...912..129Z, 2023ApJ...949...71Z}.
For the scenario invoking the large-scale kink instabilities,
a high optical polarization value of $\sim$ $20\,\%$ and a smooth polarization degree (PD)/PA modulation are predicted by \cite{2021MNRAS.501.2836B}, with a relatively low X-ray polarization expected by \cite{2021ApJ...912..129Z}.
In the weakly magnetized shock scenario,
the magnetic field in the high-energy emission region is more orderly than in the low-energy emission region, so the degree of polarization is predicted to increase with the energy band \citep{1985ApJ...298..114M, 2016MNRAS.463.3365A, Tavecchio_F-2018-Landoni_M-MNRAS.480.2872T}.

Recently, the Imaging X-ray Polarimetry Explorer (IXPE) observed the blazar Mrk~501 (\citealp{2022Natur.611..677L}) and found that its X-ray polarization ($\sim$ $10\,\%$) was higher than its optical polarization ($\sim$ $5\,\%$), which supported the shock scenario with an energy-stratified electron population.
However, the observed PDs of Mrk~501 (\citealp{2022Natur.611..677L}), together with the other blazer polarization observations (Mrk~421, \citealp{2022ApJ...938L...7D}; PG~1553+113, \cite{2023ApJ...953L..28M}; BL~Lacertae, \citealp{2023ApJ...942L..10M}), are significantly lower than the theory predictions (e.g., $\sim$ $30\,\%$, \citealp{Tavecchio_F-2018-Landoni_M-MNRAS.480.2872T}).
In \cite{Tavecchio_F-2018-Landoni_M-MNRAS.480.2872T}, the decaying behavior of the shock-generated magnetic fields (e.g., \citealp{2012SSRv..173..491S}) is considered.
However, the shock-generated magnetic fields is set to completely parallel to the shock front,
which is the same by setting a high $\zeta_{\rm sg}$ in Equation~(\ref{eq: anisotropy of sgMF}) of this paper
(also see the discussion in Section~\ref{Results:1}).
The difference between the IXPE's observations and theory predictions could come from the fact that the turbulence in the emission region will reduce the anisotropy of the shock magnetic field.
In this paper,
we studied the polarization properties of the shock-generated magnetic fields (sgMFs) at different degrees of anisotropy.

The Very Long Baseline Array (VLBA) carried out radio polarization imaging of the nearest and brightest blazar Mrk 421 (\citealp{1991rc3..book.....D}), and found that the observed PA direction was almost perpendicular to the jet-axis on the sub-pc scale \citep{2014A&A...571A..54L, 2013EPJWC..6107004L}, which means that there is a large-scale radial magnetic field.
Interestingly, the large scale radial magnetic field was also found in the supernova remnant \citep{1996ApJ...472..245J, 2008ApJ...678..939Z, 2013ApJ...772L..20I, 2017ApJ...849L..22W}, for example, the recent X-ray polarization imaging of Cassiopeia A by IXPE revealed the existence of radial magnetic fields \citep{2022ApJ...938...40V}.
\cite{2013ApJ...772L..20I} proposed that the large-scale radial magnetic field of the supernova remnant could be driven by the Richtmyer-Meshkov instability (RMI),
in which the RMI grows once the incident shock strikes the corrugated contact discontinuity separating two fluids of different densities \citep{richtmyer1960taylor, meshkov1969instability}.
In blazars, the RMI could grow gradually when the shock sweeps through the emission region with non-uniform density distributions, and then the radial (parallel to shock normal) magnetic field component could be amplified from the background magnetic field (\citealp{Sano_T-2012-Nishihara_K-ApJ.758.126S}).
In this paper, we introduce the radial large-scale magnetic fields (rmMFs) generated by the RMI in the shock model.

The present paper is structured as follows.
In Section \ref{Models}, we present our model and basic calculation method.
In Section \ref{Results}, we present the polarization behavior of the internal-shock emission for the blazar case.
In Section \ref{Discussion}, the discussions about the internal-shocks in blazars are discussed based on the IXPE Observations.
Finally, we summary the conclusions in Section~\ref{Conclusions}.

\section{Model}\label{Models}
In the internal-shock model,
the central engine of blazars is intermittently ejecting shells of relativistic
plasma at varying speeds.
These shells subsequently collide
and a pair of shocks traveling in opposite directions in the frame of the shocked fluid are formed.
As the shocks propagate in the unshocked part of shells,
they convert the ordered bulk kinetic
energy of the plasma into the magnetic field energy
and random kinetic energy of the particles.
At and near the shock front, the electrons are accelerated and
the random magnetic field is formed.
While flowing downstream,
electrons produce synchrotron emission within the magnetic field in situ
and inverse-Compton radiation,
where the synchrotron emission and the inverse-Compton radiation are
responsible for the low-frequency (radio-optical/UV) emission and
high-frequency (X-ray-$\gamma$-ray) emission from blazars, respectively.
In this paper, we revisit the polarimetric features of the internal-shocks
by considering a more general distribution of magnetic fields.
To simplify,
we study the situation that
a shock is formed at the collision radius $R_{\rm is}=3\times10^{16}\,\rm{cm}$
and propagates in a cone-shaped jet shell with thickness $R_{\rm th}=R_{\rm is}/\Gamma^2$
and opening angle $\Theta_{\rm j}$.
Hereafter, the superscript prime is used to denote the quantities measured in the co-moving frame of the downstream
and $t'$ is used to represent the time measured in the rest frame of the downstream since the collision of two jet shells.
Since both the electrons and the magnetic fields are advected in the post-shock region
and the corresponding energy decays with time,
the subscript ``$_{...,0}$'' is used to denote the quantities associated with those generated at time $t'_0$.
For example, if an electron is injected at the time $t'_0$,
the Lorentz factor $\gamma'_{{\rm e}}$ of this electron at time $t'$ is represented with $\gamma'_{{\rm e},0}(t')$.
Here, the distance of this electron relative to the shock front is $(t'-t'_0)\upsilon'_{\rm sh}$ with $\upsilon'_{\rm sh}\approx c$ being the speed of the shock front in the comoving frame of the jet flow.
The radiation at a moment $t'$ should be the sum of the radiation from the electrons injected at different $t'_0\in[0,t']$, i.e., an integral of the radiation
over the volume of the post-shock region.

\subsection{Coordinates used in the Model}\label{Sec_coordinates}
We first define three right-handed Cartesian coordinates, i.e., $\hat{1}\hat{2}k$-coordinate, $x'y'\beta'$-coordinate, and $\hat{1}'\hat{2}'k'$-coordinate. The related axes are defined as follows.

\begin{itemize}
  \item
   The unit vector $\vec{e}_{k}$ ($\vec{e}_{k'}$) of the $k$-axis ($k'$-axis) is along the line of sight and points to the observer in the rest frame of the blazar (co-moving frame of the local jet flow).
   Then, the $\vec{e}_{k}$ and $\vec{e}_{k'}$ have the following relation:
  \begin{equation}\label{eq:coordinate C}
  \vec{e}_{k'}=\mathcal{D}\{\vec{e}_{k}+\vec{e}_{\beta}[(\Gamma-1) \vec{e}_{\beta} \cdot \vec{e}_{k}-\Gamma \beta]\},
  \end{equation}
  where $\vec{\beta}c=\beta c\vec{e}_{\beta}$ is the velocity of the local jet flow in the rest frame of the blazar,
  $c$ is the light speed, $\mathcal{D}=1 /[\Gamma(1-\vec{\beta}\cdot \vec{e}_{k})]$ is the Doppler factor of the jet
  flow relative to the observer,
  and $\Gamma=1/\sqrt{1-\beta^2}$ is the bulk Lorentz factor of the local jet flow.
  \item
  Taking $\vec{e}_{\text{jet-axis}}$  as the unit vector of the jet axis moving toward us,
  the unit vectors of the $\hat{1}$-axis and $\hat{2}$-axis are defined as
  \begin{equation}\label{eq:coordinate A}
  \vec{e}_{\hat{2}}=\frac{\vec{e}_{\text{jet-axis}}\times \vec{e}_{k}}{\left|\vec{e}_{\text{jet-axis}} \times \vec{e}_{k}\right|},\;
  \quad \vec{e}_{\hat{1}}=\vec{e}_{\hat{2}}\times \vec{e}_{k}.
  \end{equation}
  The $\hat{1}$-axis and $\hat{2}$-axis of the $\hat{1}\hat{2}k$-coordinate are used to describe the observed polarization direction of the radiation.
  In the $\hat{1}\hat{2}k$-coordinate,
  we introduce a spherical coordinate with polar angle $\Theta$ and azimuthal angle $\Phi$
  to describe the emission region and the polarization angle.
  Here, the $\Theta=0$ is along the direction of $\vec{e}_{k}$
  and $\Phi$ is the azimuthal angle in $\hat{1}\hat{2}$-plane from the $\hat{1}$-axis.
  By setting the viewing angle of the jet-axis as $\Theta_{\rm v}$,
  one can have $\vec{e}_{\text{jet-axis}}=(-\cos\Theta_{\rm v}, 0, \sin\Theta_{\rm v})$.
  \item
  The co-moving frame $x'y'\beta'$-coordinate is used to describe the structure of the magnetic fields in the local jet flow.
  In this coordinate, the unit vector $\vec{e}_{\beta'}$ of the $\beta'$-axis is along the direction of the local jet flow and the unit vectors of $x'$-axis and $y'$-axis are as follows:
  \begin{equation}
  {{\vec e}_{y'}} = {{{{\vec e}_{\beta'} } \times {{\vec e}_{k'}}} \over {|{{\vec e}_{\beta'} } \times {{\vec e}_{k'}}|}},\;\quad \vec{e}_{x'} = \vec{e}_{y'} \times {\vec{e}_{\beta'} }.
  \end{equation}
  In order to describe the direction distribution of the magnetic fields in $x'y'\beta'$-coordinate,
  we introduce a spherical coordinate with polar angle $\theta'$ and azimuthal angle $\varphi'$.
  The polar axis of this spherical coordinate is along the $\beta'$-axis,
  and $\varphi'$ is the the azimuthal angle in the $x'y'$-plane from the $x'$-axis.
  \item
  In the co-moving frame of the jet flow, it is useful to define the $\hat{1'}\hat{2'}k'$-coordinate to describe the polarimetric properties of the radiation.
  The unit vectors of the $\hat{2'}$-axis and $\hat{1'}$-axis are respectively defined as
  \begin{equation}
  \quad \vec{e}_{\rm 2'}=\vec{e}_{y'},
  \quad \vec{e}_{\rm 1'} = \vec{e}_{2'}\times \vec{e}_{k'}.
  \end{equation}
  In our calculations, the polarization of the radiation is first
  decomposed into the direction of $\vec{e}_{\hat{1}'}$ and $\vec{e}_{\hat{2}'}$
  and then converted into the vector in the $\hat{1}\hat{2}k$-coordinate.
  The details can be found in \cite{2018ApJ...860...44L}.
  In the $\hat{1'}\hat{2'}k'$-coordinate,
  we also introduce a spherical coordinate with polar angle $\theta'_B$ and azimuthal angle $\varphi'_B$ to describe the direction of the magnetic field.
  Here, the $\theta'_B$ would be the pinch angle of the emitting electron in the magnetic field,
  and $\varphi'_B$ is the azimuthal angle of the magnetic field in $\hat{1'}\hat{2'}$-plane from the
  $\hat{1'}$-axis and thus is related to the polarization angle.
  The values of $\theta'_B$ and $\varphi'_B$ are used in estimating the synchrotron emission of electrons, e.g., Equations~(\ref{f_nu})-(\ref{u_nu}).
  
\end{itemize}

\subsection{Magnetic Fields in the Downstream }\label{S1}
The synchrotron emission of the internal-shock is related to three types of magnetic field:
the ordered background magnetic fields (obMFs) $\vec{B}'_{\rm ob}$,
the shock-generated random magnetic fields (sgMFs) $\vec{B}'_{\rm sg}$,
and the magnetic fields generated by the Richmyer-Meshkov instability (rmMFs) $\vec{B}^{\prime}_{\rm rm}$.

\emph{Ordered background magnetic fields.\;}
The ordered background magnetic fields in the emission region is carried from the central engine of blazars.
The geometric configuration of $\vec{B}'_{\rm ob}$ is usually set to toroidal or radial structures in the emission region.
In general, the strength of the ordered background magnetic fields
is lower than that of the sgMFs at very small distance from the shock,
but is stronger than that of the sgMFs at very high distance from the shock.
However, the emission polarization of the internal-shocks with only strong obMFs has been extensively studied
(e.g., \citealp{2014ApJ...789...66Z,2016ApJ...817...63Z}).
In addition, the strength of the rmMFs would be strong compared with
that of the ordered background magnetic fields in the post-shock region
(e.g., \citealp{2012ApJ...758..126S, 2013ApJ...772L..20I}).
This paper is focused on the emission polarization of the internal-shock
with anisotropic distribution sgMFs and rmMFs.
To simplify the problem, the obMFs is set as
$B'_{\rm ob}= f_{\rm{ob}} B'_{\rm {sh,0}}$
with $f_{\rm{ob}}=10^{-3}$,
where $B'_{\rm {sh,0}}$ is the initial strength of the shock-generated magnetic field.

\emph{Shock-generated random magnetic fields.\;}
In the internal-shock model, the random magnetic field is generated by the electric current filaments generated from Weibel instability (\citealp{Weibel_ES-1959-PhRvL.2.83W}).
Some numerical simulations show that the filaments only
survive within a microscopic scale behind the shock front,
and then they would interact with each other and force coalescence in the downstream region (\citealp{Silva_LO-2003-Fonseca_RA-ApJ.596L.121S, Medvedev_MV-2005-Fiore_M-ApJ.618L.75M, Chang_P-2008-Spitkovsky_A-ApJ.674.378C}).
The simulations reveal that the small scale turbulent magnetic field generated by the shock decays with time as a power-law form when it flows away from the shock front to the downstream (\citealp{chang2008long,lemoine2013synchrotron,lemoine2013magnetization}).
Then, the strength evolution of the sgMFs formed at time $t'_0$ can be described as
\begin{eqnarray}\label{eq:B_sh}
B'_{\rm sg,0}(t')  = \left\{ {\begin{array}{*{20}{c}}
{0,}&{t' < {t'_0}}\\
{B'_{\rm sh, 0}{{\left( {1 + \frac{{t' - {t'_0}}}{t'_{\rm B}}} \right)}^{ - {\alpha _{\rm{B}}}}},}&\,{t'\geqslant t'_0}
\end{array}} \right.,
\end{eqnarray}
where $B'_{\rm sh,0}$ is the initial strength of the sgMFs generated at time $t'_0$,
$t'$ is the time since the collision of two jet shells,
$t'_{\rm B}$ is the decay timescale of the shock-generated random magnetic field,
and $\alpha_{\rm B}$ is a power-law decay index.
The effect of the sgMFs decay-behavior on the jet's emission is related to the value of $t'_{\rm B}/t'_{\rm c}$, where $t'_{\rm c}=6\pi m_{\rm e} c/(\sigma_{\rm T} \gamma'_{\rm m} {B'_{\rm sh,0}})$ is the electron synchrotron cooling timescale in the magnetic field of $B'_{\rm sh,0}$
and $\gamma'_{\rm m}$ is the minimum Lorentz factor of the shock-accelerated electrons (\citealp{Zhao_X-2014-Li_Z-ApJ.780.12Z}).
Then, it is useful to associate $t'_{\rm B}$ with $t'_{\rm c}$,
i.e., $t'_{\rm B}=\tau_{\rm B}t'_{\rm c}$ with $\tau_{\rm B}$ as a free parameter.
In general, the initial strength of the sgMFs $B'_{\rm sh,0}$ is
associated with the dissipated energy density $\varepsilon'$ of the shock as
\begin{eqnarray}\label{eq:internal energy}
B_{\mathrm{sh}, 0}^{\prime}=\sqrt{8 \pi \epsilon_{\mathrm{B}} \varepsilon'},
\end{eqnarray}
where $\epsilon_{\rm B}$ is the fraction of the shock dissipated energy used to form the sgMFs.
With the dissipation efficiency $\epsilon_{\rm dis}$,
we simply set $\varepsilon'=\epsilon_{\rm dis}L_{\rm jet}/(4\pi R_{\rm is}^{2} c \Gamma^{2} )$ for simplicity,
where $L_{\rm{jet}}$ is the kinetic energy of the jet and $\Gamma = 10$ is the Lorentz factor of the emission region.
Assuming the fraction of the dissipated energy used to accelerate electrons is $\epsilon_{\rm e}$,
one can have $\epsilon_{\rm dis}\epsilon_{\rm e}L_{\rm k} = L_{\rm obs}$ in the fast cooling case, where $L_{\rm obs}$ is the observed luminosity.
To simplifying our discussion, the values of $\epsilon_{\rm dis}=5\%$ (\citealp{Kobayashi_S-1997-Piran_T-ApJ.490.92K}), $\epsilon_{\rm e}$ = 0.3, and $L_{\rm jet}=10^{46} \rm{erg\cdot s^{-1}}$ are used in our calculations.

Simulations reveal that the sgMFs at the shock front is predominantly orthogonal to the shock normal (\citealp{Tavecchio_F-2018-Landoni_M-MNRAS.480.2872T}).
In \cite{Tavecchio_F-2018-Landoni_M-MNRAS.480.2872T}, the sgMFs are set to completely parallel to the shock front.
However, the expected polarization in such kind of magnetic field morphology is around $\sim 30\,\%$
(e.g., \citealp{Tavecchio_F-2018-Landoni_M-MNRAS.480.2872T}),
which is significantly higher than the observations.
In order to reconcile the theoretical expectations with the observations,
we revisit the polarimetric features of the internal-shock model
by considering a more general distribution of the sgMFs' direction.
Since the direction of the sgMFs is predominantly orthogonal to the shock normal,
the distribution of the sgMFs direction is described as
\begin{eqnarray}\label{eq: anisotropy of sgMF}
w'_{\rm sg}(\theta', \varphi') =(\sin\theta')^{\zeta_{\rm sg}},
\end{eqnarray}
where ${\zeta_{\rm sg}}$ is a power-law index  and $\theta'$ is the angle between the direction of sgMF and $\vec{e}_{\beta'}$ (see Section~\ref{Sec_coordinates}).
The distribution of the sgMFs' direction is assumed to remain the same during the decaying of the sgMFs and is the same for sgMFs generated at different $t'_0$.
Equation~(\ref{eq: anisotropy of sgMF}) reveals that the sgMFs with
$\vec B'_{\rm{sg},0} = B'_{\rm{sg},0}(\sin \theta '\cos \varphi ',\;{\rm{sin}}\theta '\sin \varphi ',\;{\rm{cos}}\theta ')$ takes a proportion of ${w'_{{\rm{sg}}}}(\theta ',\varphi ')d\Omega '/\!\int\!\!\!\int\!{{w'_{{\rm{sg}}}}(\theta ',\varphi ')d\Omega '} $ for the sgMFs generated at time $t'_0$,
where $d\Omega ' = {\rm{sin}}\theta 'd\theta 'd\varphi'$.

\emph{Magnetic fields generated by the Richmyer-Meshkov instability.\;}
Except the sgMFs, the magnetic fields can be amplified through the macroscopic turbulence dynamo
excited by the so-called Richtmyer-Meshkov instability.
This instability is an inevitable outcome of interactions between shock and ambient density fluctuations
(Richtmyer 1960; Meshkov 1969).
Simulations show that the magnetic field
can grow by at least two orders of magnitude compared to the magnetic energy immediately
behind the shock (i.e., obMFs),
provided the kinetic energy of turbulence injected by the Richtmyer-Meshkov instability is greater than the magnetic energy
(e.g., \citealp{Inoue_T-2011-Asano_K-ApJ.734.77I,Sano_T-2012-Nishihara_K-ApJ.758.126S, 2013ApJ...772L..20I}).
Then, we take $B_{\rm{\rm rm},0}^{\prime}=f_{\rm{rm}} B'_{\rm {sh,0}}$
with $f_{\rm{rm}}=2\times10^{-2}$, and the intensity of rmMFs in the post-shock region remains constant.
The simulations of the rmMFs is finally dominated by the radial component (e.g., \citealp{Sano_T-2012-Nishihara_K-ApJ.758.126S}).
Then, the distribution of the rmMFs' direction is described as
\begin{eqnarray}\label{eq: anisotropy of rmMF}
w'_{\rm rm}\left(\theta', \varphi'\right) =
(\cos\theta')^{\zeta_{\rm rm}},
\end{eqnarray}
where ${\zeta_{\rm rm}}$ is the anisotropy index of the rmMFs.
Equation~(\ref{eq: anisotropy of rmMF}) reveals that the rmMF with $\vec{B}'_{\rm rm,0}={B'_{{\rm{rm},0}}}(\sin \theta '\cos \varphi ',\;{\rm{sin}}\theta '\sin \varphi ',\;{\rm{cos}}\theta ')$
takes a proportion of
${w'_{{\rm{rm}}}}(\theta ',\varphi ')d\Omega '/\!\int\!\!\!\int\!{{w'_{{\rm{rm}}}}(\theta ',\varphi ')d\Omega '}$ for the rmMFs in the emission region.

\emph{Magnetic field in any region.\;}
The total magnetic field in a emission region can be expressed as
\begin{equation}\label{Eq:total magnetic}
\vec{B}'_{\rm tot,0}(t') =\left\{
\begin{array}{ll}
\vec{B}'_{\rm sg,0}(t'),&{\rm{for}}\;{\rm{only}}\;{\rm{sgMFs}}\\
\vec{B}'_{\rm sg,0}(t')+\vec{B}'_{\rm rm,0},&{{\rm{for}}\;{\rm{only}}\;{\rm{sgMFs + rmMFs,}}},\\
\vec{B}'_{\rm sg,0}(t')+\vec{B}'_{\rm ob},&{\rm{for}}\;{\rm{only}}\;{\rm{sgMFs + obMFs}}.
\end{array}\right.
\end{equation}

\subsection{Electrons prescription and Evolution}
The calculation of the evolution of the electron energy spectrum is taken from the work of \cite{2023ApJ...957..109Z}.
Again we write it as follows.
It is generally assumed that the accelerated electrons near the shock front obey a power-law distribution,
i.e., $n'_{\rm{e,0}}=N'_{\rm{e,0}}[{\gamma'_{\rm{e,0}}(t'_0)}]^{-p}$
for $\gamma'_{\rm m}<\gamma'_{{\rm e},0}(t'_0)<\gamma'_{\rm max}$ and $n'_{{\rm e},0}=0$ for others,
where ${\gamma'_{\rm{e},0}(t')}$ is the Lorentz factor of electrons at time $t'$
for these injected at the time $t'_0$,
{the normalization factor $N'_{\rm{e,0}} = L_{\rm{jet}}/[\Gamma^2 m_{\rm p}c^2(4\pi R^2_{\rm is}c\beta)]/\int [{\gamma'_{\rm{e,0}}(t'_0)}]^{-p} d\gamma'_{\rm{e,0}}(t'_0)\sim 7\times 10^{-3}$}$\rm cm^{-3}\cdot s^{-1}$ is simply set,
$p$ is the power-law index and is taken as 4.7 in our calculation,
and
$\gamma'_{\rm max}=[9 m_{\rm e}^{2}c^{4}/(8{B'_{\rm {tot,0}}}(t'_0)q_{\rm e}^{3})]^{1/2}$ is the maximum Lorentz factor of the shock-accelerated electrons,
and $m_{\rm p}$, $m_{\rm e}$, and $q_{\rm e}$ are the rest mass of the proton,
the rest mass of the electron, and the charge of the electron, respectively.
 In our scenario, both the electrons and the magnetic fields are advected in the post-shock region
and the corresponding energy decays with time.
During the advection in the post-shock region,
the electrons suffer from the radiation cooling by the synchrotron radiation and inverse-Compton radiation, i.e.,
\begin{equation}\label{Eq:gamma_e_coolin}
\frac{d\gamma'_{{\rm e,0}}(t'
)}{dt'}=-\frac{\sigma_{\rm T}}{6 \pi m_{\rm e} c}[{\gamma'_{{\rm e,0}}(t'
)}]^{2} {B'}^{2}_{\rm tot,0}(t')\cdot(1+Y),
\end{equation}
where $Y$ is the Compton parameter.
Generally, the radiation cooling of the electrons is strong compared
with that due to the expansion of the jet shell in the internal-shocks.
Then, the adiabatic cooling of electrons is neglected in Equation~(\ref{Eq:gamma_e_coolin}).
Since the magnetic field decays with time, the value of $Y$ can be described as
\begin{equation}\label{Eq:Y}
Y_0=\tilde{Y}[{B'_{\rm tot,0}}(t'_0)/{B'_{\rm tot,0}(t')}]^{2},
\end{equation}
where $\tilde{Y}$ denotes the ratio of inverse-Compton radiation to synchrotron radiation power at the shock front
and $\tilde{Y}=0.5$ is set in this paper (\citealp{Zhao_X-2014-Li_Z-ApJ.780.12Z}).
With Equations~(\ref{Eq:gamma_e_coolin}) and (\ref{Eq:Y}),
the Lorentz factor of an electron varies with time as
\begin{equation}\label{Eq:gamma_e_evolution}
\frac{1}{\gamma'_{{\rm e,0}}(t')}=\left\{\begin{array}{ll}
\frac{1}{\gamma'_{{\rm e,0}}(t'_0)}-\frac{\sigma_{\rm T}}{6 \pi m_{\rm e} c}\left[-s(t'-t'_0)+\frac{t'_{\rm B} B'^{2}_{\rm sh,0}}{1-2 \alpha_{\rm B}}\left(1-T^{-2 \alpha_{\rm B}+1}\right)\right], & \alpha_{\rm B} \neq 1 / 2,\\
\\
\frac{1}{\gamma'_{{\rm e,0}}(t'_0)}-\frac{\sigma_{\rm T}}{6 \pi m_{\rm e} c}\left[-s(t'-t'_0)-t'_{\rm B} B'^{2}_{\rm sh,0} \ln T\right], &
\alpha_{\rm B}=1 / 2,
\end{array}\right.
\end{equation}
where
$s={B'}^{2}_{\rm ob}+{B'}^{2}_{\rm rm,0}+\tilde{Y} [{B'}_{\rm tot,0}(t'_0)]^{2}$
and $T=1+(t'-t'_0)/t'_{\rm B}$.
Equation~(\ref{Eq:gamma_e_evolution}) reveals that the evolution of $\gamma'_{{\rm e,0}}(t')$ depends on
$s$, $B'_{\rm sh,0}$, $t'_{\rm B}$, and $\alpha_{\rm B}$.
The effect of $\tilde{Y}$ on $\gamma'_{{\rm e,0}}(t')$ and thus the synchrotron radiation spectrum is made
through the factor of $s=B'^{2}_{\rm ob }+B'^{2}_{\rm rm,0}+\tilde{Y} [B'_{\rm tot,0}(t'_0)]^{2}$.
It reveals that the effects of $\tilde{Y}$ and $B'_{\rm tot,0}(t'_0)$ on the synchrotron radiation spectrum are the same if both $B'_{\rm sh,0}$ and $B'_{\rm ob}+B'_{\rm rm,0}$ remain the same in Equation~(\ref{Eq:gamma_e_evolution}).
That is to say the effect of $\tilde{Y}$ on the synchrotron radiation spectrum can be reflected by changing the value of $B'_{\rm tot,0}(t'_0)$.
This behavior has been partially showed in the figure~3 and discussed in the section~4 of \cite{Zhao_X-2014-Li_Z-ApJ.780.12Z}.
In this paper , $\tilde{Y}=0.5$ is set for simplifying our discussion and fittings.

In general,
the energy distribution of electrons is obtained based on the standard continuity equation
(e.g. \citealp{1999MNRAS.306..551C,2021ApJ...911...13L}).
However, one can obtain the energy distribution of electrons based on the following process.
In our scenario, if an electron is injected at the time $t'_0$ with Lorentz factor $\gamma'_{\rm{e},0}(t'_0)$,
the Lorentz factor of this electron would decay to $\gamma'_{{\rm e},0}(t')$ at time $t'$.
Correspondingly, the total number of electrons in the $[\gamma'_{e,0}(t'),\gamma'_{e,0}(t')+d\gamma'_{e,0}(t')]$
is $n'_{\rm{e}, 0}d \gamma'_{\rm{e}, 0}(t'_0)$,
which is used in Equations~(\ref{f_nu})-(\ref{u_nu}).
That is to say, the energy distribution of the electrons at time $t'$ for the electrons injected at the time $t'_0$ can be described as $n'_{\rm{e}, 0}d \gamma'_{\rm{e}, 0}(t'_0)/d \gamma'_{\rm{e}, 0}(t')$.

\subsection{The radiation and polarization}\label{S3}
The detailed process for estimating the emission polarization of the synchrotron emission has been presented
in previous works (e.g., \citealp{2006A&A...453..621D, 2009ApJ...698.1042T, 2018ApJ...860...44L}).
The calculations of the synchrotron emission in this paper are the same and presented as follows.

\emph{Synchrotron emission for electrons injected at the time $t'_0$.\;\,}
 In our scenario, both the electrons and the magnetic fields are advected in the post-shock region.
During the advection, the electrons injected at time $t'_0$ are cooled to the Lorentz factor of $\gamma'_{{\rm e,0}}(t')$ at time $t'$.
With the magnetic field $\vec B'_{\rm tot,0}(t')$ in situ,
the synchrotron emission of the electrons with $\gamma'_{{\rm e,0}}(t')$
in the $\hat{1}'\hat{2}'k'$-coordinate can be described as
\begin{equation}\label{f_nu}
f_{\nu,0}^{\prime} = B^{\prime}_{\rm{tot,0}}(t')\sin{\theta'_{\rm B}}\int_{\gamma'_{\rm m}}^{\gamma'_{\rm max}}F(x)n_{\mathrm{e}, 0}^{\prime}d \gamma_{\mathrm{e,0}}^{\prime}(t'_0),
\end{equation}
\begin{equation}\label{q_nu}
q_{\nu,0}^{\prime}= -\frac{\sqrt{3} {q_{\rm e}}^{3}}{m_{e} c^{2}}B^{\prime}_{\rm{tot,0}}(t')\sin{\theta'_{\rm B}}\cos(2\varphi'_{\rm B})\int_{\gamma'_{\rm m}}^{\gamma'_{\rm max}}G(x)n_{\mathrm{e}, 0}^{\prime}d \gamma_{\mathrm{e,0}}^{\prime}(t'_0),
\end{equation}
\begin{equation}\label{u_nu}
u_{\nu,0}^{\prime}= -\frac{\sqrt{3} {q_{\rm e}}^{3}}{m_{e} c^{2}}B^{\prime}_{\rm{tot,0}}(t')\sin{\theta'_{\rm B}}\sin(2\varphi'_{\rm B})\int_{\gamma'_{\rm m}}^{\gamma'_{\rm max}}G(x)n_{\mathrm{e}, 0}^{\prime}d \gamma_{\mathrm{e,0}}^{\prime}(t'_0),
\end{equation}
where 
$\theta'_B$ is the pinch angle of the emitting electron in the magnetic field,
i.e., $\sin\theta'_B=|\vec B'_{\rm tot,0}(t')\times \vec{e}_{k'}|/|\vec B'_{\rm tot,0}(t')|$,
$\varphi'_B$ is the azimuthal angle of the magnetic field $\vec B'_{\rm tot,0}(t')$ in $\hat{1'}\hat{2'}$-plane from the $\hat{1'}$-axis,
$x=\nu(1+z)/(\mathcal{D}\nu'_{\rm ch})$ with $\nu$ being the observed frequency, $\nu'_{\rm ch}={q_{\rm e}}B'_{\rm{tot},0}\left(t^{\prime}\right)\sin\theta'_{\rm B}[\gamma'_{{\rm e},0}(t')]^2/(2\pi m_{\rm e}c)$ is the characteristic frequency of the photon emitted by the electron with Lorentz factor $\gamma'_{e,0}(t')$,
$F(x)=x\int_{x}^{\infty}K_{5/3}(\xi)d\xi$, $G(x)=xK_{2/3}(x)$, and $K_{5/3}(x)$ and $K_{2/3}(x)$ are the modified Bessel function of $5/3$ and $2/3$ order, respectively.
One should note that if an electron is injected at time $t'_0$,
the Lorentz factor of this electron is $\gamma'_{e,0}(t')$ at time $t'$. Since only the energy of the electrons decays with time $t'$, the total number of the electron with Lorentz factor in the $[\gamma'_{e,0}(t'),\gamma'_{e,0}(t')+d\gamma'_{e,0}(t')]$ can be described as $n'_{\rm{e}, 0}d \gamma'_{\rm{e}, 0}(t'_0)$, which is used in Equations~(\ref{f_nu})-(\ref{u_nu}).
Since the distribution of the sgMFs and rmMFs,
the synchrotron emission from a region can be described as
\begin{equation}
\overline{f'_{\nu,0}}=\iint\!\!\iint f'_{\nu,0} w'_{\rm sg}\left(\theta_{\rm sg}^{\prime}, \varphi_{\rm sg}^{\prime}\right)w'_{\rm rm}\left(\theta_{\rm rm}^{\prime}, \varphi_{\rm rm}^{\prime}\right) d \Omega_{\rm sg}^{\prime} d \Omega_{\rm rm}^{\prime},
\end{equation}
\begin{equation}
\overline{q'_{\nu,0}}=\iint\!\!\iint q'_{\nu,0} w'_{\rm sg}\left(\theta_{\rm sg}^{\prime}, \varphi_{\rm sg}^{\prime}\right)w'_{\rm rm}\left(\theta_{\rm rm}^{\prime}, \varphi_{\rm rm}^{\prime}\right) d \Omega_{\rm sg}^{\prime} d \Omega_{\rm rm}^{\prime},
\end{equation}
\begin{equation}
\overline{u'_{\nu,0}}=\iint\!\!\iint u'_{\nu,0} w'_{\rm sg}\left(\theta_{\rm sg}^{\prime}, \varphi_{\rm sg}^{\prime}\right)w'_{\rm rm}\left(\theta_{\rm rm}^{\prime}, \varphi_{\rm rm}^{\prime}\right) d \Omega_{\rm sg}^{\prime} d \Omega_{\rm rm}^{\prime}.
\end{equation}
Based on the above results, the polarization degree and polarization angle in the co-moving frame can be read as
\begin{equation}
\Pi'_{\nu,0}=\frac{\sqrt{\overline{q'_{\nu,0}}^2+\overline{u'_{\nu,0}}^2}}{\overline{f'_{\nu,0}}},\;\;
\cos(2 \chi'_{\nu,0})=\frac{\overline{q'_{\nu,0}}}{\sqrt{\overline{q'_{\nu,0}}^2+\overline{u'_{\nu,0}}^2}},\;\;
\sin(2 \chi'_{\nu,0})=\frac{\overline{u'_{\nu,0}}}{\sqrt{\overline{q'_{\nu,0}}^2+\overline{u'_{\nu,0}}^2}}.
\end{equation}
Correspondingly, the Stokes parameters in the the observer frame are
\begin{equation}
f_{\nu,0}=\overline{f'_{\nu,0}}\mathcal{D}^{3},\;\;
\Pi_{\nu,0}=\Pi'_{\nu,0},\;\;
q_{\nu,0}  = {f_{\nu,0}}{\Pi_{\nu,0} }\cos(2{\chi_{\nu,0}}),\;\;
u_{\nu,0}=f_{\nu,0} \Pi_{\nu,0} \sin(2\chi_{\nu,0}),
\end{equation}
where $\cos (2\chi_{\nu,0}) =({{1 - {{\tan }^2}{\chi_{\nu,0}}}})/({{1 + {{\tan }^2}\chi _{\nu,0}}})$,
$\sin (2{\chi_{\nu,0}}) = {{2\tan {\chi _{\nu,0}}}}/({{1 + {{\tan }^2}{\chi _{\nu,0}}}})$,
$\tan\chi_{\nu,0} = ({{\sin \Phi \tan {\chi'_{\nu,0}} - \cos \Phi }})/({{\sin \Phi + \cos \Phi \tan {\chi'_{\nu,0}}}})$, and  $\Phi$ is the azimuthal angle of the emission region in $\hat{1}\hat{2}k$-coordinate
(see Section~\ref{Sec_coordinates}).
For the details about the relation of $\chi _{\nu,0}$ and $\chi'_{\nu,0}$,
please see the equation~5 in \cite{2019ApJ...870...96L}.

\emph{Synchrotron emission for electrons in the post-shock region.\;\,}
The total synchrotron emission is from the entire post-shock region.
Considering the equal arrival time surface effects,
the observed synchrotron radiation flux density and the Stokes parameters are described as
\begin{align}\label{EQ:Observation_bin}
F_{\rm obs}(\nu, t_{\rm obs})&=\frac{1+z}{4\pi D^2_L}
\int_{0}^{\min(t',t'_{\rm end})}{\!\!\int\!\!\!\!\int\limits_{\rm EATS}\!\!\!\!{{f_{\nu,0}d\Omega}}\;dt'_0},\\
Q_{\rm obs}(\nu, t_{\rm obs})&=\frac{1+z}{4\pi D^2_L}
\int_{0}^{\min(t',t'_{\rm end})}{\!\!\int\!\!\!\!\int\limits_{\rm EATS}\!\!\!\!{q_{\nu,0}d\Omega}\;dt'_0},\\
U_{\rm obs}(\nu, t_{\rm obs})&=\frac{1+z}{4\pi D^2_L}
\int_{0}^{\min(t',t'_{\rm end})}{\!\!\int\!\!\!\!\int\limits_{\rm EATS}\!\!\!\!{u_{\nu,0}d\Omega}}\;dt'_0,\label{EQ:Observation_end}
\end{align}
where $D_{\rm L}$ is the luminosity distance,
$t'_{\rm end}=R_{\rm th}\Gamma/{\upsilon '_{{\rm{sh}}}}$,
$d\Omega=\sin\Theta d\Theta d\Phi$ is the solid angle of the emission region,
and ``EATS'' represents the integration over the equal-arrival time surface (EATS).
For the jet flow shocked at time $t'_0$,
the observed time of a photon emitted from
$\vec{R}=R(\cos\Phi\sin\Theta, \sin\Phi\sin\Theta, \cos\Theta)$ is estimated as
\begin{equation}\label{EQ:EATS}
{t_{{\rm{obs}}}} = \left\{ {\frac{{R - {R_{\rm{0}}}}}{c} + \int_{{R_0}}^R {(1 - \beta )\frac{{dr}}{{c\beta }}}  + \frac{{R(1 - \cos \Theta )}}{c}} \right\}(1 + z),
\end{equation}
where $R_0 = R_{\rm is} - \upsilon'_{\rm sh}t'_0/\Gamma$ and $R =R_0 + (t' - t'_0)\Gamma \beta c$ are the radius of the jet flow at the collision time and $t'$, respectively.
According to Equations~(\ref{EQ:Observation_bin})-(\ref{EQ:Observation_end}),
the observed polarization degree and angle is estimated as
\begin{align}
\Pi&=\frac{\sqrt{Q^2_{\rm obs}+U^2_{\rm obs}}}{F_{\rm obs}}, \notag \\
\chi&=\frac{1}{2}\arctan\left(\frac{U_{\rm obs}}{Q_{\rm obs}}\right).
\end{align}

As previously knew, the emission polarization depends not only on the magnetic field structure
but also the observational angle $\Theta_{\rm v}$ (e.g., \citealp{Tavecchio_F-2018-Landoni_M-MNRAS.480.2872T}).
This paper is focused on the emission polarization of the internal-shock
with anisotropic distribution sgMFs and rmMFs.
In such kind of magnetic fields,
\cite{Tavecchio_F-2018-Landoni_M-MNRAS.480.2872T} has shown that
if a low viewing angle $\Theta_{\rm v}$ is set, e.g., $\Theta_{\rm v}\lesssim \Theta_{\rm j}$,
the polarization degree of the synchrotron emission would be significantly low.
In our calculations, we also find that the emission polarization from the situation with $\Theta_{\rm v}=\Theta_{\rm j}$ is significantly lower than that from the situation with $\Theta_{\rm v}=\Theta_{\rm j}+1/\Gamma$,
which is inconsistent with the IXPE's observations ($\sim 10\%$, e.g., \citealp{2022Natur.611..677L, 2022ApJ...938L...7D}).
In addition, a large part of the blazars are likely to be off-axis observations (e.g., \citealp{2008A&A...488..905G,2014A&A...571A..54L,2023ApJS..266...37A}).
Then, $\Theta_{\rm v}=\Theta_{\rm j}+1/\Gamma$ is set in our calculations.
Here, $z =0.03$ is the cosmological redshift and $\upsilon'_{\rm sh}=c
$ is set.
In order to consistent with the observed radiation spectrum,
the value of $p=4.7$ is adopted for the electron injection.

\section{Results}\label{Results}

\subsection{Polarization for the Situation with only Anisotropic sgMFs}\label{Results:1}
It was earlier found that the magnetic fields generated by shocks are not isotropic
and is dominated by the component perpendicular to the shock normal
(\citealp{2014ApJ...794...46C, 2021Galax...9...37T, 2016ApJ...817...63Z, Tavecchio_F-2018-Landoni_M-MNRAS.480.2872T}).
In this paper, we revisit the polarization of the internal-shock with such kind of the sgMFs.
Different from previous works (e.g., \citealp{Tavecchio_F-2018-Landoni_M-MNRAS.480.2872T}),
Equation~(\ref{eq: anisotropy of sgMF}) is introduced to describe the distribution of the sgMFs' direction
in this paper.
If a higher value of the anisotropy index $\zeta_{\rm sg}\,(>0)$ is adopted in Equation~(\ref{eq: anisotropy of sgMF}),
the more of the sgMFs would be perpendicular to the shock normal.
In Figure~{\MyFigA}, we show the frequency dependent $\Pi$ (upper-panel) and $\chi$ (bottom-panel),
where different values of $\zeta_{\rm sg}$,
i.e., 0.5 (orange lines), 1 (green lines), 1.5 (blue lines),
2 (red lines), and 3 (purple lines), are adopted.
In this figure,
the solid and dash lines are corresponding to different decay behavior of the sgMFs,
i.e., $\alpha_{B}=1.0$ and $0.5$ for solid and dash lines, respectively.
By comparing different colors of the solid or dash lines,
one can find that the anisotropy index $\zeta_{\rm sg}$ would significantly
affect the value of $\Pi$ for all wave band.
If a higher value of the anisotropy index $\zeta_{\rm sg}$ is adopted,
the obtained degree $\Pi$ of the polarization would be higher for the same wave band.
This behavior is not affected by adopting different decay behavior of the sgMFs,
i.e., different $\alpha_{\rm B}$.
Similar to the results found in \cite{Tavecchio_F-2018-Landoni_M-MNRAS.480.2872T},
$\alpha_{\rm B}$ mainly makes its effect on the polarization of the intermediate frequency emission (e.g., the optical band),
rather than the high frequency emission (e.g., the X-ray band) and low frequency emission (e.g., the radio band).
It is worth pointing out that for the case with $\zeta_{\rm sg}=2$ (i.e., purple lines),
the polarization of X-rays is around $\Pi\sim 25\,\%$ and that of optical band is around $\Pi\sim 12\%$,
which is similar to those of figure 4 in \cite{Tavecchio_F-2018-Landoni_M-MNRAS.480.2872T}.
It reveals that Equation~(\ref{eq: anisotropy of sgMF}) with $\zeta_{\rm sg}=2$
roughly describe the distribution of the sgMFs' direction in \cite{Tavecchio_F-2018-Landoni_M-MNRAS.480.2872T}.

In Figure~{\MyFigB}, we shows that relations of $\Pi-\zeta_{\rm sg}$ for X-ray and radio bands,
where the ``$\times$'' and ``$+$'' symbols are for X-ray and radio bands, respectively.
The dependence of the polarization on the value of anisotropy index $\zeta_{\rm sg}$ can be easily found.
In Figure~{\MyFigB}, it can be easily found that the value of $\Pi$ increases quickly with $\zeta_{\rm sg}$.
We fit the $\Pi-\zeta_{\rm sg}$ relations with a function of $\Pi=a[1-\exp(-\zeta_{\rm sg}/b)]$,
where the $a$ and $b$ are the free parameters in the fittings.
The relations of $\Pi\sim 44.5\%[1-\exp(-\zeta_{\rm sg}/2.6)]$ (red line) and
$\Pi\sim 20\%[1-\exp(-\zeta_{\rm sg}/2.4)]$ (black line) are obtained for X-ray and radio bands, respectively.
The above two relations can be understood as follows.
(1) In the situation with $\zeta_{\rm sg}\rightarrow\infty$, almost all of the sgMFs's direction is perpendicular to the shock normal in the shock co-moving frame.
This kind of case is the same as that the electrons are in the ordered magnetic fields.
Correspondingly, the polarization in the high-energy emission can be related to the spectral index $\beta$ as
$\Pi=(-2\beta+2)/(-2\beta+10/3)$ or the injected index $p$\footnote{The relation of $p-\beta$ of the high-energy synchrotron emission for electrons in a non-decaying magnetic field scenario can be described as $p=-2\beta+1$ for fast cooling situation.
However, it is worthy pointing out that the $p-\beta$ of the high-energy synchrotron emission
for electrons in a decaying magnetic field scenario
may be different from $p=-2\beta+1$ since a decay behavior of the magnetic field makes its effect on the radiation spectrum of the synchrotron emission.}
as $\Pi\sim (p+1)/(p+7/3)$.
In Figures~{\MyFigA} and {\MyFigB},
the value of $p=4.7$ is adopted for the electron injection and thus $\Pi\sim 81\,\%$.
For low frequency emission, e.g., radio band, the polarization can be described as
$\Pi=(-2\beta+2)/(-2\beta+10/3)$ with $\beta=1/3$, i.e., $\Pi=50\%$.
One should note that the geometric structure of the emission region would makes its effect on the polarization of the emission
and the asymmetric geometry of the emission region would depolarize the synchrotron emission.
In the situation with $\zeta_{\rm sg}\rightarrow\infty$,
the relations of $\Pi\sim 44.5\%[1-\exp(-\zeta_{\rm sg}/2.6)]$ and
$\Pi\sim 20\%[1-\exp(-\zeta_{\rm sg}/2.4)]$ are reduced to $\Pi\sim 44.5\%$ and $20\%$ for the X-ray and radio bands, respectively.
This reveals that the geometric structure of the emission region depolarizes the synchrotron emission.
By comparing the fitting results with the theoretical expectations,
one can have
\begin{itemize}
\item for high frequency emission (e.g., X-ray band)
\begin{equation}\label{EQ:sg_high}
\Pi  \approx \frac{100}{1.82}(- 2\beta  + 2)/(- 2\beta  + 10/3)[1 - \exp ( - {\zeta _{{\rm{sg}}}}/2.6)]\,\%,
\end{equation}
\item for low frequency emission (e.g., radio band)
\begin{equation}\label{EQ:sg_low}
\Pi  \approx \frac{100}{2.5}(- 2\beta  + 2)/(- 2\beta  + 10/3)[1 - \exp ( - {\zeta _{{\rm{sg}}}}/2.4)]\,\%,
\end{equation}
\end{itemize}
where the suppression factor of $1.82$ and $2.5$ in the denominators are related to the depolarization effect of the emission region.
Since the emission region of the high-frequency emission is small compared with that of the low-frequency emission, the depolarization effect of the emission region would be low for the high-frequency emission
compared with that for the low-frequency emission.
This is consistent that the suppression factor in Equation~(\ref{EQ:sg_high}), i.e., $1.82$,
is low than that in Equation~(\ref{EQ:sg_low}), i.e., $2.5$.

Since the sgMFs is mainly distributed along the direction perpendicular to the shock normal,
the polarization of the emission would be parallel to the shock normal for all wave band.
In the bottom panel of Figure~{\MyFigA}, the value of $\chi$ is indeed $0^\circ$ for all wave band.

\subsection{Polarization in the Situation with only sgMFs + rmMFs}\label{Sec_sg_rm}
The Richtmyer-Meshkov instability appears when an incident shock wave strikes a corrugated contact
discontinuity separating two fluids of different densities
(\citealp{richtmyer1960taylor, meshkov1969instability}).
In the blazars, there are very likely that the density in the jets are not uniformly distributed
and thus Richtmyer-Meshkov instability may be activated.
In this part, we study the emission polarization of the internal-shock with both sgMF and rmMFs.
It is reasonable to believe that the sgMFs would dominate the magnetic field around the shock front.
In this situation, the rmMFs has little effect on the polarization of the high-frequency emission
and thus $\zeta_{\rm sg}\sim 1$ is required to explain the IXPE observations (i.e., $\Pi\sim 10\%$ in the X-ray band, \citealp{2022Natur.611..677L}).
Then, $\zeta_{\rm sg}\sim 1$ is adopted in the following studies.

In this paper, the sgMFs is assumed to decay with time as Equation~(\ref{eq:B_sh})
and the rmMFs is assumed to not decay with time.
If the sgMFs generated at time $t'_0$ decays quickly, the magnetic field in situ
would be dominated by the rmMFs for significantly high $t'-t'_0$.
If not, the sgMFs would dominate the magnetic field all the way.
Here, the decay behavior of the sgMFs is reflected by varying the value of $\alpha_{\rm B}$.
In Figure~{\MyFigE}, we show the relations of $\nu-\Pi$ and $\nu-\chi$
in the situation that the rmMFs dominates the magnetic field quickly.
Here the value of $\alpha_{\rm B} = 1.0$ is adopted.
In this figure, the anisotropy index of the rmMFs $\zeta_{\rm rm}=0.25$ (green line), 0.5 (blue line),
0.75 (red line), and 1 (black line) are adopted.
One can find that there is a transition of the $\chi$ from $90^\circ$ to $0^\circ$ at transition frequency $\sim10^{15}-10^{16}$~Hz
based on the bottom panel of Figure~{\MyFigE}.
It reveals that the magnetic field of the low-frequency emission is dominated by the rmMFs
and that of the high-frequency emission is dominated by the sgMFs.
This is owing to that the sgMFs and the rmMFs are mainly distributed along the direction perpendicular
and parallel to the shock normal, respectively.
In this situation, the emission polarization in the emission region with magnetic field dominated by
the sgMFs or rmMFs would be parallel ($\chi=0^\circ$) or perpendicular ($\chi=90^\circ$) to the shock normal.
It is worth pointing out that the $\Pi-\nu$ relation around the transition frequency is peculiar
(see also the green solid line in Figure~{\MyFigD}).
As $\nu$ decreases,
the polarization degree $\Pi$ decreases from $\sim12.5\%$ at the high-frequency emission to
$\sim0\%$ at around the transition frequency,
and subsequently increases to a certain value (e.g., $\sim7.5\%$ for the black line)
at the low-frequency emission.
The value of $\Pi\sim0\%$ around the transition frequency reveals that the magnetic field
responsible for the transition frequency emission is almost isotropic.

In Figure~{\MyFigF}, we shows that relations of $\Pi-\zeta_{\rm rm}$ for optical and radio bands in the situations the same as those of Figure~{\MyFigE} but with different $\zeta_{\rm rm}$,
where the ``$\times$'' and ``$+$'' symbols represents the numerical data for the optical and radio bands, respectively.
By fitting the $\Pi-\zeta_{\rm rm}$ relations with a function of $\Pi=a[1-\exp(-\zeta_{\rm sg}/b)]$,
the relations of $\Pi\sim 28\%[1-\exp(-\zeta_{\rm sg}/2.7)]$ (red line) and
$\Pi\sim 23\%[1-\exp(-\zeta_{\rm sg}/2.4)]$ (black line) are obtained for the optical and radio bands, respectively.
Considering the effect of geometric structure of the emission region on the emission polarization,
we can have
\begin{itemize}
\item for optical band
\begin{equation}\label{EQ:rm_high}
\Pi  \approx\frac{100}{2.8}(- 2\beta  + 2)/(- 2\beta  + 10/3)[1 - \exp ( - {\zeta _{{\rm{rm}}}}/2.7)]\,\%,
\end{equation}
\item for radio band
\begin{equation}\label{EQ:rm_low}
\Pi  \approx \frac{100}{2.2}(- 2\beta  + 2)/(- 2\beta  + 10/3)[1 - \exp ( - {\zeta _{{\rm{rm}}}}/2.4)]\,\%,
\end{equation}
\end{itemize}
where the suppression factor of $2.8$ and $2.2$ in the denominators are related to the depolarization effect of the emission region.

If the sgMFs does not decay quickly, the sgMFs would dominate the magnetic field in most of the emission region.
In Figure~{\MyFigG}, we shows the emission polarization in the situation with $\alpha_{\rm B} = 0.5$,
where the anisotropic index $\zeta_{\rm rm}$ = 0 (black line), 1 (red line), 2 (blue line), 3 (green line),
and 4 (orange line) are adopted.
It can be easily found that the value of $\chi$ remains $0^\circ$ for all wave band
and thus the emission polarization is mainly subject to the distribution of the sgMFs.
Compared to the case with only sgMFs, e.g., the green dashed line in Figure~{\MyFigA},
the polarization degree $\Pi$ in the radio band is low for all the cases in in Figure~{\MyFigE}.
In addition, the value of $\Pi$ decreases as the $\zeta_{\rm rm}$ increases according to Figure~{\MyFigG}.
This behavior reveals that
the exit of the rmMFs in the emission region weakens the anisotropy of the magnetic field distribution,
even through the rmMFs is not dominated the magnetic field in the main body of the emission region.
Taking the case with $\zeta_{\rm rm}=0$ as an example,
the polarization degree $\Pi\sim5.2\%$ is obtained for the radio band,
which is lower than that of the case with only sgMFs, i.e., $\Pi\sim6.8\%$ (the green dashed line in Figure~{\MyFigA}).
This indicates that the fraction of emission from the electrons in sgMFs to those in the rmMFs can be expressed as
$f_{\rm sg} = 5.2\%/6.8\%=77\%$
and
$f_{\rm rm} = 1-f_{\rm sg}=23\%$, respectively.
One should note that the main distribution direction of the rmMFs is perpendicular to the main distribution direction of the sgMFs if $\zeta_{\rm rm}$ is not equal to zero.
Then, the depolarization on the radio emission of the jet
would be stronger if a higher $\zeta_{\rm rm}$ is adopted.
We find that the total polarization degree of radio band can be roughly described as
$\Pi_{\rm sg}(\zeta_{\rm sg})f_{\rm sg}-\Pi_{\rm rm}(\zeta_{\rm rm})\times f_{\rm rm}$,
where $f_{\rm sg}$ ( $f_{\rm rm}=1-f_{\rm sg}$ ) is
the contribution fraction of the electrons in the sgMFs (rmMFs) for the radio emission
and can be estimated based on the case with $\zeta_{\rm rm}=0$,
$\Pi_{\rm sg}(\zeta_{\rm sg})$ is estimated based on
Equation~(\ref{EQ:sg_low}) for a given $\zeta_{\rm sg}$,
and $\Pi_{\rm rm}(\zeta_{\rm rm})$ is estimated based on
Equation~(\ref{EQ:rm_low}) for a given $\zeta_{\rm rm}$.

\subsection{Effects of the obMFs or viewing angle on the Polarization}
In general, the strength of the ordered background magnetic fields
is lower than that of the sgMFs at very small distance from the shock,
but is stronger than that of the sgMFs at very larger distance from the shock.
In addition, the strength of the rmMFs would be strong compared with
that of the ordered background magnetic fields in the post-shock region
(e.g., \citealp{2012ApJ...758..126S, 2013ApJ...772L..20I}).
Since the emission polarization of the internal-shocks with only strong obMFs has been extensively studied
(e.g., \citealp{2014ApJ...789...66Z,2016ApJ...817...63Z}),
we study the emission polarization of the internal shocks with only sgMFs+obMFs,
of which the results are shown in Figure~{\MyFigD}.
Here, the situation with fast decaying sgMFs, i.e., $\alpha_{\rm B} =2$, is adopted,
and the $\zeta_{\rm sg}=1$ is set.
In this figure, different kind of obMFs' morphology, i.e., toroidal obMFs (black solid line)
and radial obMFs (blue solid line) are discussed,
and the emission polarization of the internal shocks
with only sgMFs is also plotted with green solid line for comparisons.
From this figure,
one can find that the polarization at the low-frequency emission
is very different from that from the situation with only sgMFs.
It reveals that the obMFs becomes dominant in much larger post-shock region,
where produces the low-frequency emission.
Then, it is easy to see that the polarization degree at the low-frequency emission is very high,
e.g., 20-40 \%, which is similar to that with only obMFs (e.g., \citealp{2014ApJ...789...66Z,2016ApJ...817...63Z,
Tavecchio_F-2018-Landoni_M-MNRAS.480.2872T}).
Interesting, there is a dip in the $\Pi-\nu$ relation and a transition in the $\chi-\nu$ relation
at $\nu\sim 10^{15}-10^{16}$~Hz for the situation with radial obMFs.
This is similar to those found in Figure~{\MyFigE} (see the discussion in Section~\ref{Sec_sg_rm}).
In addition, a bump appears below the dip in the $\Pi-\nu$ relation at $\nu\sim 10^{14}$~Hz.
This is owing to that (1) for the internal shock emission in the situation with only obMFs,
the $\Pi$ remains almost constant in $\nu<10^{12}$~Hz
and increases with rising $\nu$ in $\nu\gtrsim10^{12}$~Hz;
(2) the high-frequency emission is mainly from the the electrons in the sgMFs dominated region, i.e., around the shock front.
Moreover, the value of $\Pi$ in this situation is significantly lower than
that in the situation with only obMFs.
Then, the combination of (1) and (2) leads to the formation of a bump at $\nu\sim 10^{14}$~Hz.

We also study the situation with slow decaying sgMFs, e.g., $\alpha_{\rm B} =0.5$,
of which the polarization of the internal shock emission is similar to that in Figure~{\MyFigA}
and thus is not showed in this paper.
In addition, the strength of the rmMFs would be strong compared with
that of the ordered background magnetic fields in the post-shock region
(e.g., \citealp{2012ApJ...758..126S, 2013ApJ...772L..20I}).
However, if the strength of the rmMFs is weaken than that of obMFs,
the polarization of the internal-shock emission is similar to those in Figure~{\MyFigD} or Figure~{\MyFigA}.
Then, the situation with sgMFs + rmMFs + obMFs is not shown in this paper.

The emission polarization depends not only on the magnetic field structure
but also the viewing angle $\Theta_{\rm v}$ (e.g., \citealp{Tavecchio_F-2018-Landoni_M-MNRAS.480.2872T}).
For the internal-shock with anisotropic distribution sgMFs and rmMFs,
\cite{Tavecchio_F-2018-Landoni_M-MNRAS.480.2872T} has shown that
if a low viewing angle $\Theta_{\rm v}$ is set,
the polarization degree of the synchrotron emission would be significantly low.
In Figure~{\MyFigD}, we also show the polarization of the internal shock emission
in the situation with $\Theta_{\rm v}=\Theta_{\rm j}$ (dashed lines).
A very low polarization degree is indeed found from this figure.
It implies that if the internal-shock scenario is applied to explain the IXPE's observations ($\sim 10\%$, e.g., \citealp{2022Natur.611..677L, 2022ApJ...938L...7D}),
a high viewing angle is required.
Indeed, a large part of the blazars are likely to be off-axis observations (e.g., \citealp{2008A&A...488..905G,2014A&A...571A..54L,2023ApJS..266...37A}).

\section{Discussion based on the IXPE Observations}\label{Discussion}
Recently, the IXPE, combined with other telescopes, performed multi-wavelength observations of
the blazars Mrk~501, Mrk~421, and PG~1553+113.
The blazars Mrk~501, Mrk~421 and PG~1553+113 are classified as high-synchrotron-peaked BL~Lacertae objects ( $\nu_{\rm{peak}}>10^{15}\,\rm{Hz}$, \citealp{2022Natur.611..677L, 2022ApJ...938L...7D, 2023ApJ...953L..28M}),
of which the X-ray emission is dominated by the synchrotron radiation of the electrons.
The IXPE's observations may reveal that the particle acceleration in shocks may operate in the blazar jet
(\citealp{2022Natur.611..677L, 2022ApJ...938L...7D}).
In this section, we apply the IXPE's observations associated with the radio/optical observations to estimate the properties of the sgMFs and rmMFs.
\begin{itemize}
\item
\cite{2022Natur.611..677L} performed two observations on the Mrk~501 with IXPE,
accompanied by observations across the electromagnetic spectrum from multiple observatories.
The results from these observations (\citealp{2022Natur.611..677L}) are summarized as follows:
(1) The X-ray linear polarization degree $\Pi_{\rm X}$ is of around 10$\%$,
the host-galaxy corrected-intrinsic optical polarization degree $\Pi_{\rm O}$ is $\sim5\,\%$,
and the radio polarization degree $\Pi_{\rm R}$ is $ \sim1.5\,\%$.
(2) The polarization of the radio-to-X-ray bands are all aligned with the jet axis within uncertainties.
(3) There is no evidence of polarization variability during either IXPE observation.
In Section~\ref{Results},
we have found that the polarization of the jet emission in the internal-shock scenario is along the jet axis
if the sgMFs dominate the magnetic field in most of the emission region.
Then, the sgMFs would dominate the magnetic field in most of the emission region during the IXPE observations of \cite{2022Natur.611..677L}.
In the internal-shock scenario for the blazar jet's emission, the X-rays are mainly from the shock front,
in which the magnetic field is dominated by the sgMFs.
Then, based on Figure~{\MyFigB}, we can have $\zeta_{\rm sg}\sim 0.7$ for
the shock responsible to the X-rays in the Mrk~501.
Correspondingly, the polarization degree of the radio (optical) band
would be around $5\%$ ($\gtrsim 5\%$) based on Figure~{\MyFigB} if the rmMFs can be neglected compared with sgMFs.
However, the observation of $\Pi_{\rm R}\sim1.5\,\%$ is very low compared with the expectation of Figure~{\MyFigB}.
It may reveal that emission of the electrons in the rmMFs would make a significant contribution
to the radio band and  the obMFs can be neglected.
The contribution fraction $f_{\rm rm}$ of the emission for the electrons in the rmMFs
and the anisotropy index $\zeta_{\rm rm}$ have the following relation:
$\Pi_{\rm sg}(\zeta_{\rm sg})f_{\rm sg}-\Pi_{\rm rm}(\zeta_{\rm rm})\times f_{\rm rm}=1.5\%$,
where $f_{\rm rm}=1-f_{\rm sg}$, $\Pi_{\rm sg}(\zeta_{\rm sg})=\Pi_{\rm sg}(\zeta_{\rm sg}=0.7)=5\%$,
and $\Pi_{\rm rm}(\zeta_{\rm rm})$ is estimated based on
Equation~(\ref{EQ:rm_low}) for a given $\zeta_{\rm rm}$.
Taking $\zeta_{\rm rm}=0$ as an example,
one has $\Pi_{\rm rm}(\zeta_{\rm rm})=0$, $f_{\rm sg}=1.5\%/5\%=30\%$, and $f_{\rm rm}=70\%$.
The $\zeta_{\rm rm}$-dependent $f_{\rm rm}$ is showed in Figure~{\MyFigH} with black line,
where the value of $f_{\rm rm}$ varies from 70\% to 16\% for different $\zeta_{\rm rm}$.
It is worth pointing out that most of the rmMFs are along the direction of the shock normal for
the case with $\zeta_{\rm rm}\gtrsim 3$.

\item
The results from the first observation of Mrk 421 are similar to the observation of Mrk 501, i.e.,
$\Pi_{\rm X} \sim 15\,\%$, $\Pi_{\rm O} \sim 2.7\,\%$, and $\Pi_{\rm R} \sim 3\,\%$ (\citealp{2022ApJ...938L...7D}).
Importantly, the polarization angle of radio-X-ray bands is greatly deviated from the direction of the jet axis.
However, the Mrk~421 jet has a wide opening angle of ~ $60^\circ$, and the jet may bend tens of degrees on a small scale (\citealp{2022ApJ...938L...7D}).
Thus, \citealp{2022ApJ...938L...7D} proposed that the polarization angle of radio-X-ray bands is roughly along the direction of the jet.
In this scenario,
one can have $\zeta_{\rm sg} = 1.5$ for the sgMFs based on Figure~{\MyFigB}.
Correspondingly, the theoretical polarization degree of the radio band in the situation with only sgMFs
is $\Pi_{\rm R} \sim 9\,\%$, which is significantly different from the observations.
It also reveals that the emission of the electrons in the rmMFs would make a significant contribution to the radio band and  the obMFs can be neglected.
Based on the relation of $\Pi_{\rm sg}(\zeta_{\rm sg})f_{\rm sg}-\Pi_{\rm rm}(\zeta_{\rm rm})\times f_{\rm rm}=3\%$ and $\Pi_{\rm sg}(\zeta_{\rm sg})=9\%$,
one can have $f_{\rm sg}=3\%/9\%=30\%$ and $f_{\rm rm}=70\%$ for the case with $\zeta_{\rm rm}=0$.
For other value of $\zeta_{\rm rm}>0$, the value of $f_{\rm rm}\sim 20\%-70\%$ can be found
in Figure~{\MyFigH} with red line.
It should be noted that the optical polarization is lower than the radio polarization,
which seems to contradict the predictions of the shock-accelerated energy-stratified electron model.
This is more likely that the host-galaxy make a significant contribution in the optical band,
which has found in the observation of the Mrk~501 (Fig.\,3 in \citealp{2022Natur.611..677L}).

It should be noted that the IXPE conducted three observations of Mrk~421 (\citealp{2022ApJ...938L...7D, 2023NatAs...7.1245D, 2023arXiv231210732A})
and the latter two observations reported a large-angle rotation of the polarization angle only in the X-ray band
rather than in optical and radio bands (\citealp{2023NatAs...7.1245D,2023arXiv231210732A}).
This may imply that the X-ray emission region is different from the optical/radio emission region in these two observations.

\item
\cite{2023ApJ...953L..28M} reveals an orphan optical polarization swing of PG~1553+113 during IXPE observation.
IXPE performed the observations of PG~1553+113 on 2023 February 1-2 and 2023 February 7-8.
The time average degree of X-ray polarization at both observed periods is $\Pi_{\rm X} \sim 10.1\,\%$
with $\chi_{\rm X}\sim 86^{\circ}$.
During the observation period of February 1-2,
the optical polarization degree $\Pi_{\rm O} \sim 2.2\,\%$ with a significant rotation of the polarization angle is found.
However, there is no significant variety in the X-ray polarization angle.
This implies that the emission region of the optical band
may be different from that of the X-ray band in this period.
Then, we do not discuss the observation results in this period.
During the observation period of February 7-8,
the optical polarization $\Pi_{\rm O} \sim 4.2\,\%$ and radio polarization $\Pi_{\rm R} \sim 3\,\%$
are found,
and the polarization angle of the radio-X-ray bands is roughly consistent.
The observed X-ray polarization direction is found to be oblique to the parsec-scale jet direction with a difference of $\sim45^\circ$,
which is similar to that observed for Mrk~421 during its first IXPE observation in 2022 May.
Since the local direction of the jet may be different from the parsec-scale jet direction,
\cite{2022ApJ...938L...7D} proposed the polarization direction of X-ray band
may be in the same direction of the local jet around the emission region.
With Equation~(\ref{EQ:sg_high}),
one can have $\zeta_{\rm sg}=0.7$ for the sgMFs
based on the observed X-ray polarization degree $\Pi_{\rm X} \sim 10.1\,\%$.
Correspondingly, the polarization degree of the radio band is estimated to be $\Pi_{\rm R} \sim 5\,\%$
for the case with only sgMFs.
It also reveals that the emission of the electrons in the rmMFs would make a significant contribution to the radio band and the obMFs can be neglected.
By comparing with the observed $\Pi_{\rm R} \sim 3\,\%$,
the contribution fraction of the emission from the electrons in the rmMFs
is $f_{\rm rm}\sim 10\%-40\%$,
where the $\zeta_{\rm rm}$-dependent $f_{\rm rm}$ is showed in Figure~{\MyFigH}) with blue line.

\end{itemize}
In summary, the anisotropy index of the sgMFs is around $\zeta_{\rm sg}\sim 0.7-1.5$ based on the polarization degree of the X-ray emission in blazars.
Correspondingly, one can estimate the polarization degree of the radio band
expected in the situation with only sgMFs,
which is found to be generally higher than the observed radio polarization degree in the radio band.
This reveals the emission of the electrons in the rmMFs make a significant contribution in the low-frequency emission, and the contribution fraction is around $40\%-70\%$ in the situation with $\zeta_{\rm rm}=0$.
The $\zeta_{\rm rm}$-dependent $f_{\rm rm}$ is showed in Figure~{\MyFigH} for the observations discussed above.
If most of the rmMFs are along the shock normal (i.e., $\zeta_{\rm rm}\gtrsim 3$),
the contribution fraction $f_{\rm rm}$ is around $10\%-23\%$.
We would like to point out that most of the rmMFs are along the direction of the shock normal for
the situation with $\zeta_{\rm rm}\gtrsim 3$.

\section{Conclusions}\label{Conclusions}
The morphology of the magnetic field in the jet of blazars is of great significance for the study of the radiation mechanism and particle acceleration process in situ.
The operation of the IXPE polarization detector provides an unprecedented opportunity to directly study the magnetic field morphology in the X-ray emission region.
Recent IXPE observations reveals that the X-ray polarization is higher than the optical and radio polarization
and parallel to the shock normal.
Thus, the X-ray emission of the blazars is suggested to support the shock scenario with energy-stratified electron population (\citealp{2022Natur.611..677L, 2022ApJ...938L...7D}).
In order to reconcile the theoretical expectations with observations,
we revisit the emission polarization of the internal-shock by
considering different kind of direction distribution
for the sgMFs and rmMFs,
while also considering the effect of the obMFs.

The sgMFs are mainly along the direction perpendicular to the shock normal.
Then, we introduce the $w'_{\rm sg}\propto(\sin\theta')^{\zeta_{\rm sg}}$ with $\theta'=0$ being along the shock normal to describe the direction distribution of the sgMFs in the shock co-moving frame.
Here, most of the sgMFs are distributed along the shock front in the situation with $\zeta_{\rm sg}\lesssim 3$.
In the cases with only sgMFs,
it is found that the polarization in the X-ray and radio emission
for a general jet in blazars
can be described as
$\Pi_{\rm sg}\sim 44.5[1-\exp(-\zeta_{\rm sg}/2.6)]\%$ and $\Pi_{\rm sg}\sim 20[1-\exp(-\zeta_{\rm sg}/2.4)]\%$, respectively.
The polarization degree of the optical band is between that of the X-ray band and that of the radio band.
The polarization directions of radio-X-ray bands are the same.
In addition, the above results do not affected by the decay behavior of the sgMFs,
i.e., different $\alpha_{\rm B}$ is adopted to describe the decay of the sgMFs.
In the internal-shock scenario for the blazar jet's emission,
the X-rays are mainly from the shock front,
in which the magnetic field is dominated by the sgMFs.
Based on the results of the IXPE observations, the value of $\zeta_{\rm sg}\sim 1-1.5$
is required in order to reconcile the X-ray observations.

The magnetic fields generated by the Richmyer-Meshkov instability (rmMFs),
which is triggered once the incident shock strikes the corrugated contact discontinuity separating two fluids of different densities (\citealp{richtmyer1960taylor, meshkov1969instability}),
is supposed to present in the jets.
Simulations reveal that the rmMFs are mainly distributed along the direction of the shock normal
(\citealp{2012ApJ...758..126S, 2013ApJ...772L..20I})
and thus the $w'_{\rm rm}\propto(\cos \theta')^{\zeta_{\rm rm}}$
is introduced to describe the direction distribution of rmMFs.
In the situation with $\zeta_{\rm rm}\lesssim 3$,
most of the rmMFs are distributed along the shock normal.
In the situation with sgMFs and rmMFs,
the presence of rmMFs is likely to reduce the polarization degree in the
low-frequency emission if the emission of the electrons in the sgMFs dominates the emission.
If the emission of the electrons in the rmMFs dominates the low-frequency emission,
the polarization direction of the low-frequency emission
would be perpendicular to that of the high-frequency emission (i.e., X-rays).
For the situation that the low-frequency emission is dominated by the electrons in the rmMFs,
we find that the polarization degree of the radio emission can be described as
$\Pi_{\rm rm} \approx (- 2\beta  + 2)/(- 4.4\beta  + 22/3)[1 - \exp ( - {\zeta _{{\rm{rm}}}}/2.4)]\,\%$.
For the situation that the polarization directions of the radio-X-ray bands are the same,
the polarization degree of the low-frequency (e.g., radio) emission
can be described as $\Pi_{\rm sg}(\zeta_{\rm sg})f_{\rm sg}-\Pi_{\rm rm}(\zeta_{\rm rm})\times f_{\rm rm}$,
where $f_{\rm sg}$ and $f_{\rm rm}$ are the contribution fraction of the electrons in the sgMFs
and that in the rmMFs for the radio emission, respectively.
Based on the contemporaneous radio and X-ray observations,
we find the the emission of the electrons in the rmMFs make a significant contribution
in the low-frequency emission.
If the rmMFs can be described with $\zeta_{\rm rm}=0$,
the contribution fraction $f_{\rm rm}$ is around $40\%-70\%$.
If most of the rmMFs are along the shock normal (i.e., $\zeta_{\rm rm}\gtrsim 3$),
the contribution fraction $f_{\rm rm}$ is around $10\%-23\%$.

The obMFs may significantly affect the low-frequency emission polarization.
If the sgMEs decay quickly ($\alpha_{\rm B}=2.0$) and the obMFs is stronger than the rmMFs,
the polarization degree of the radio and optical bands would be significantly high, e.g., 20-40\%.
For the recent IXPE's observations on high-synchrotron-peaked BL~Lacertae objects,
the observational polarization degree in the radio band is significantly lower than that expected from the situation with only sgMFs.
This reveals that the obMFs can be neglected in the post-shock region.
Correspondingly,
if the internal-shock scenario is applied to explain the IXPE's observations,
a high viewing angle is required.

\clearpage
\begin{figure}
\centering
 \hspace*{-1.0truecm}
 \vspace*{0.0truecm}
\includegraphics[width=0.8\textwidth]{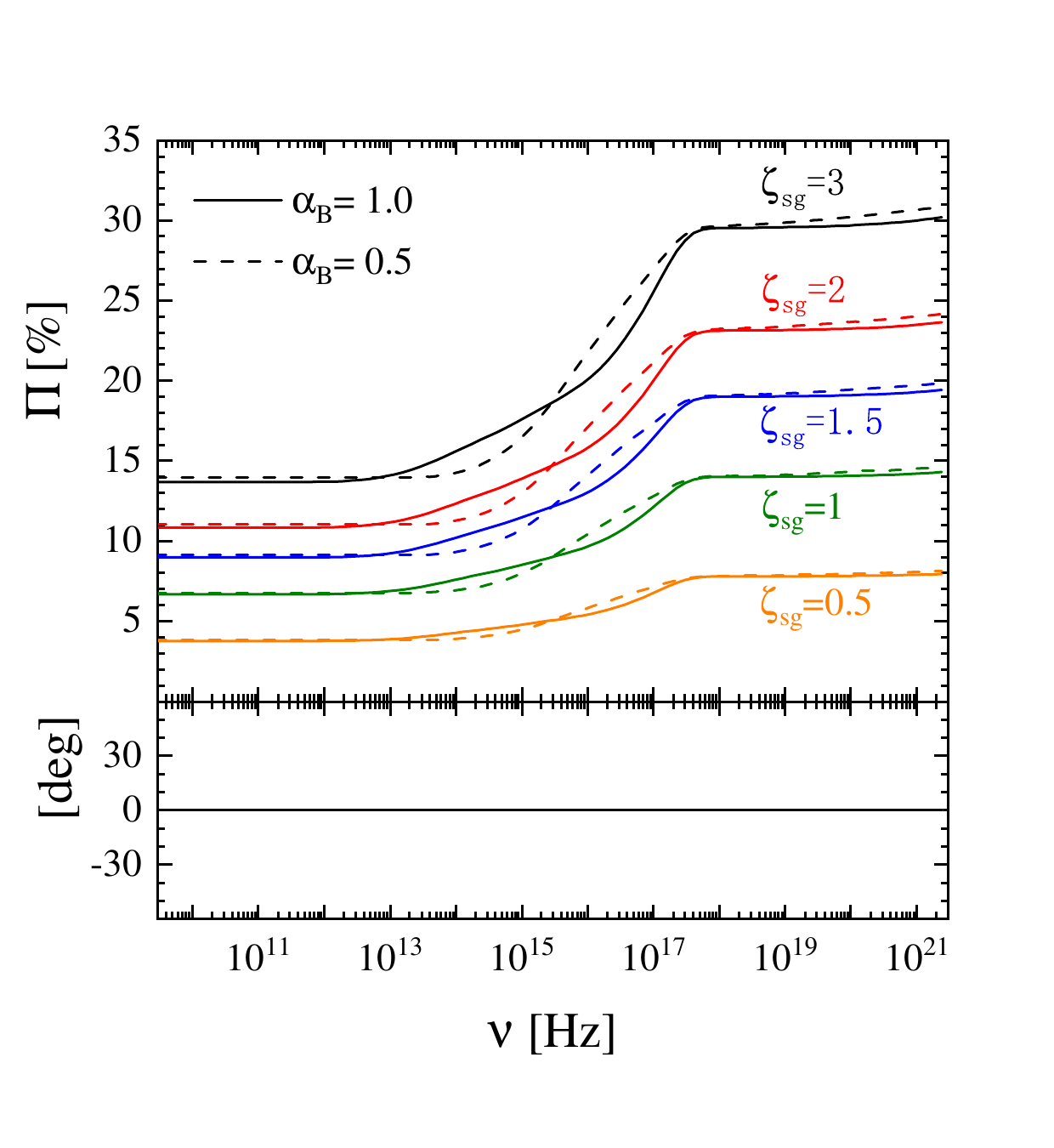}\\  
\caption{The frequency dependent $\Pi$ (upper panel) and $\chi$ (bottom panel),
where the value of $\zeta_{\rm sg}=0.5$ (orange lines), 1 (green lines), 1.5 (blue lines),
2 (red lines), and 3 (black lines) are adopted.
The solid and dashed lines are for the cases with $\alpha_{\rm B}$=1.0 and 0.5, respectively.
}
\label{MyFigA}
\end{figure}

\begin{figure}
\centering
 \hspace*{0.0truecm}
 \vspace*{0.0truecm}
\includegraphics[width=1\textwidth]{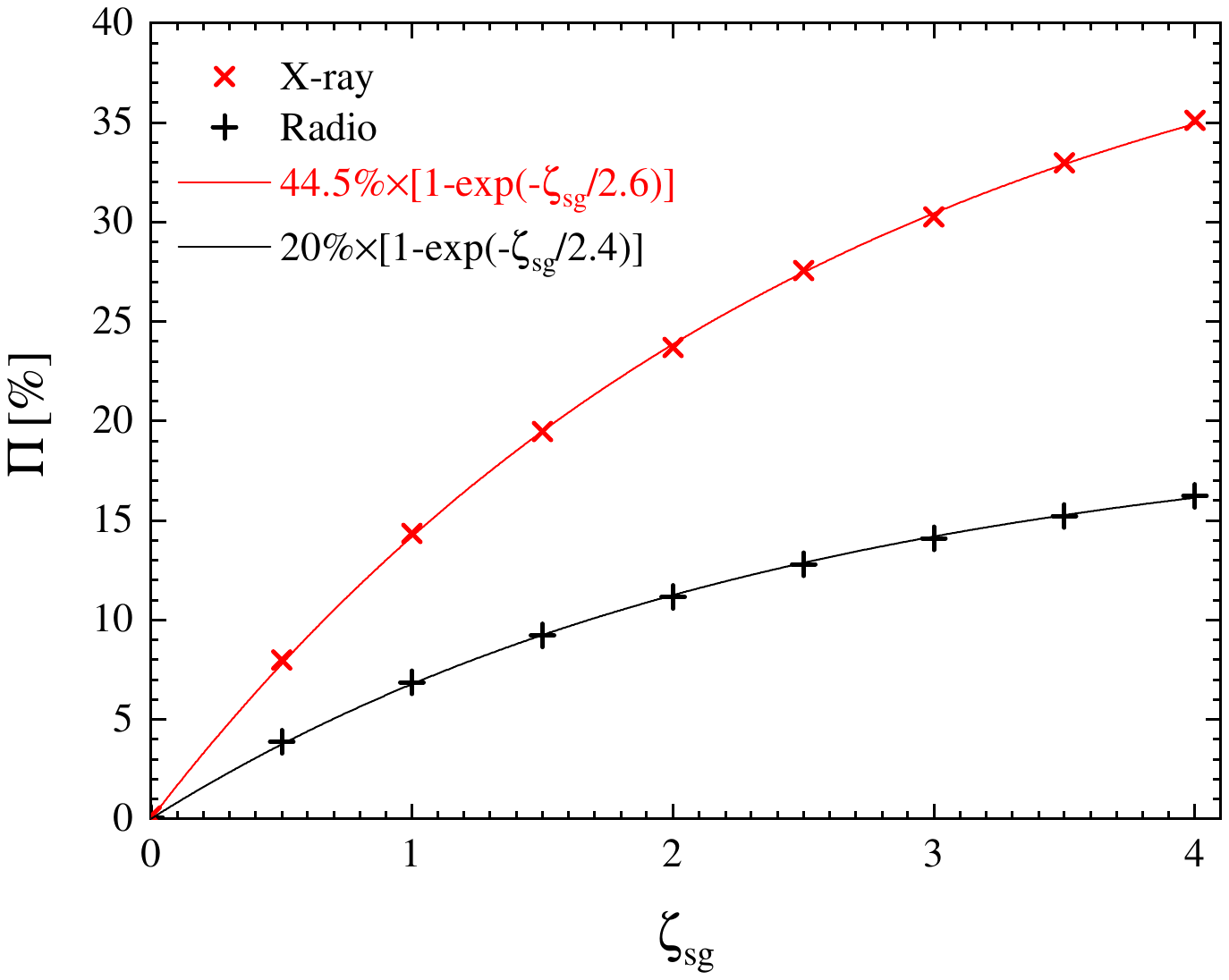}\\  
\caption{Relations of $\Pi-\zeta_{\rm sg}$ for the X-ray (red line) and radio (black line) bands
in the case with only sgMFs, where the value of $\alpha_{\rm B}$ = 1.0 is adopted.}
\label{MyFigB}
\end{figure}

\begin{figure}
\centering
 \hspace*{0.0truecm}
 \vspace*{0.0truecm}
\includegraphics[width=0.8\textwidth]{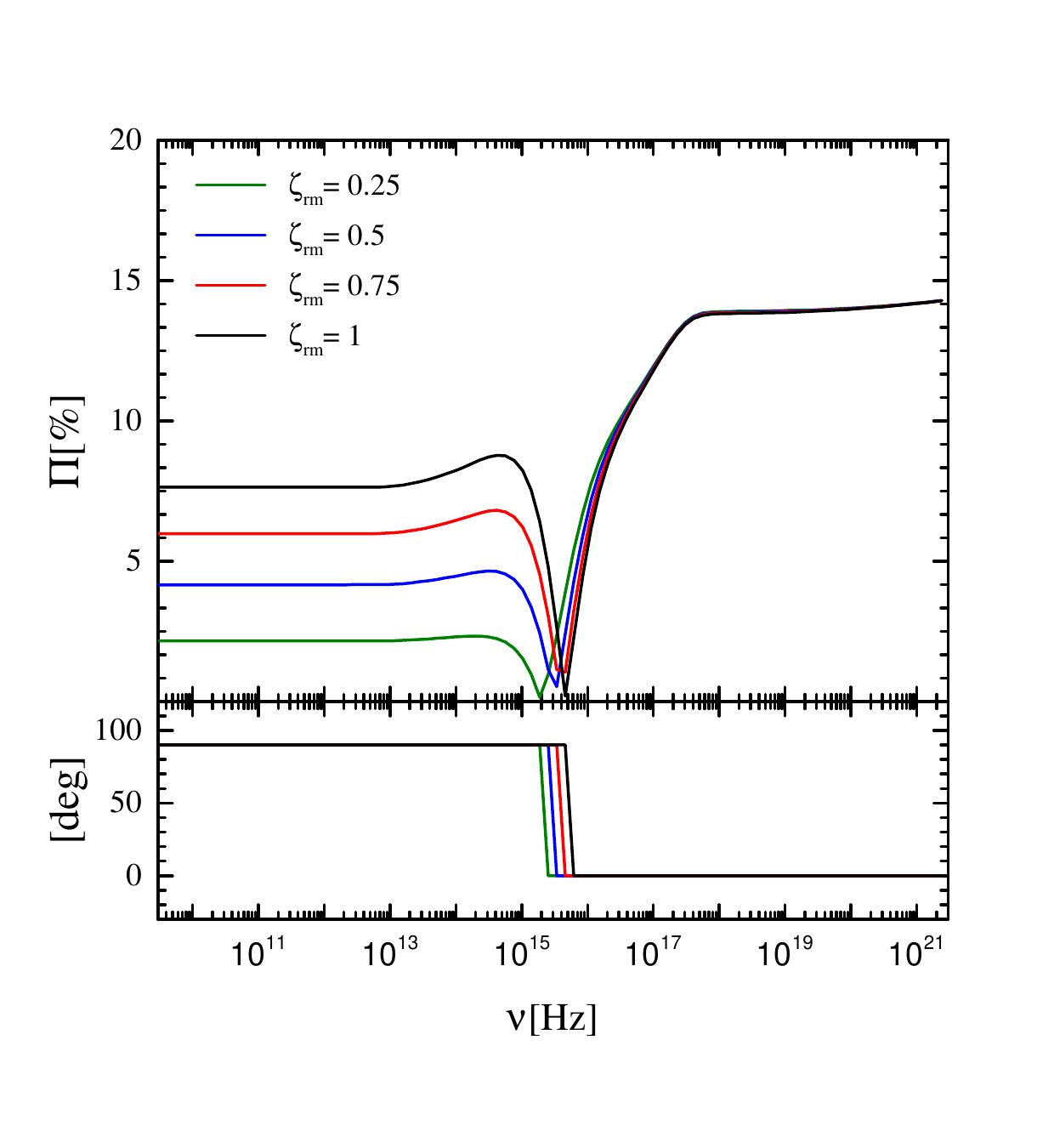}\\  
\caption{The frequency dependent $\Pi$ (upper panel) and $\chi$ (bottom panel)
from the case with both sgMFs and rmMFs,
where the $\alpha_{\rm B}$ = 1.0 and $\zeta_{\rm sg}=1$ are adopted to describe the sgMFs.
The green, blue, red and black lines correspond to the cases of $\zeta_{\rm rm}$ =0.25, 0.5, 0.75, and 1, respectively.}
\label{MyFigE}
\end{figure}

\begin{figure}
\centering
 \hspace*{0.0truecm}
 \vspace*{0.0truecm}
\includegraphics[width=0.8\textwidth]{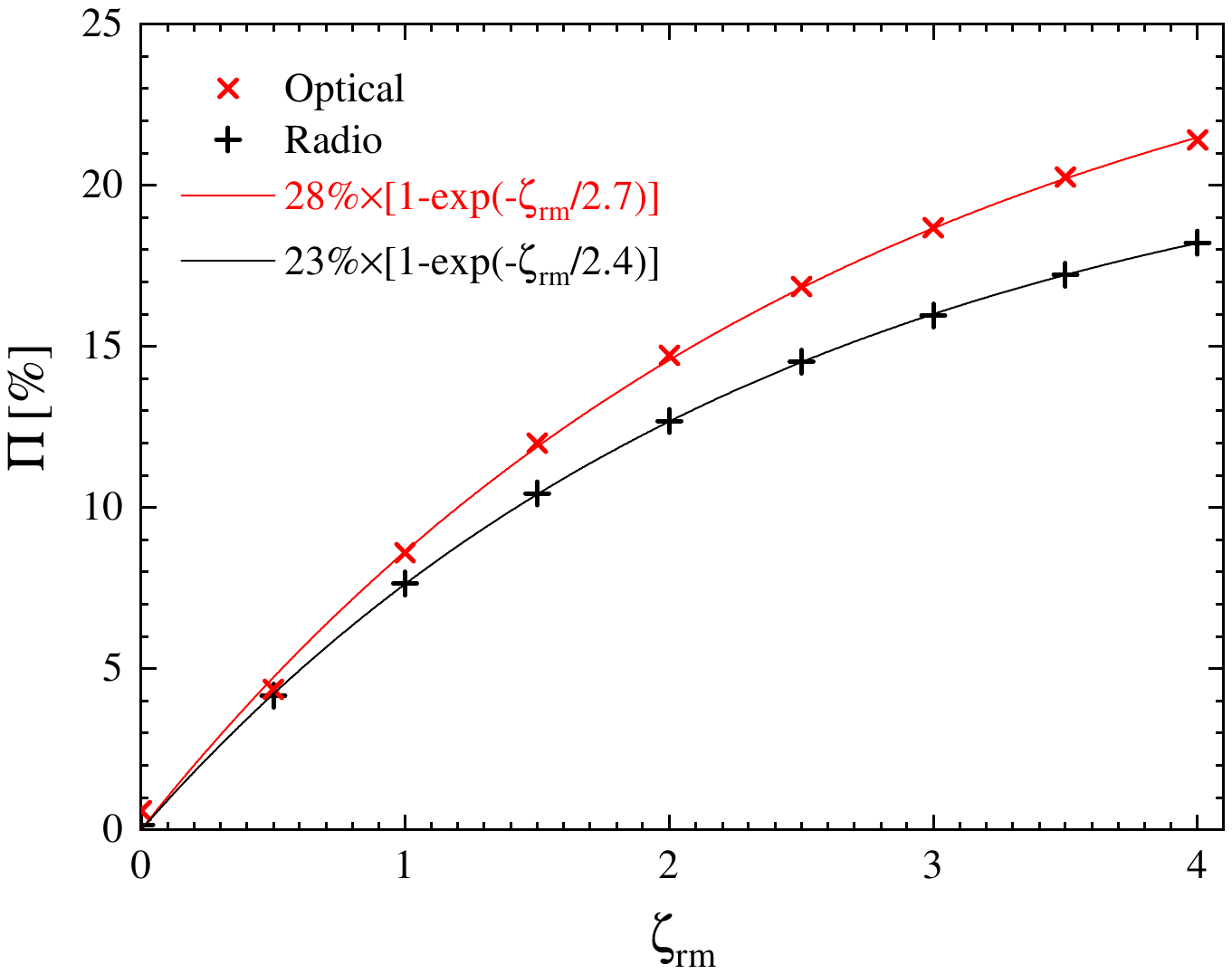}\\  
\caption{Relations of $\zeta_{\rm rm}-\Pi$ for optical (red line) and radio (black line) bands
in the case with both sgMFs and rmMFs,
where the $\alpha_{\rm B}$ = 1.0 and $\zeta_{\rm sg}=1$ are adopted to describe the sgMFs.}
\label{MyFigF}
\end{figure}

\begin{figure}
\centering
 \hspace*{0.0truecm}
 \vspace*{0.0truecm}
\includegraphics[width=0.8\textwidth]{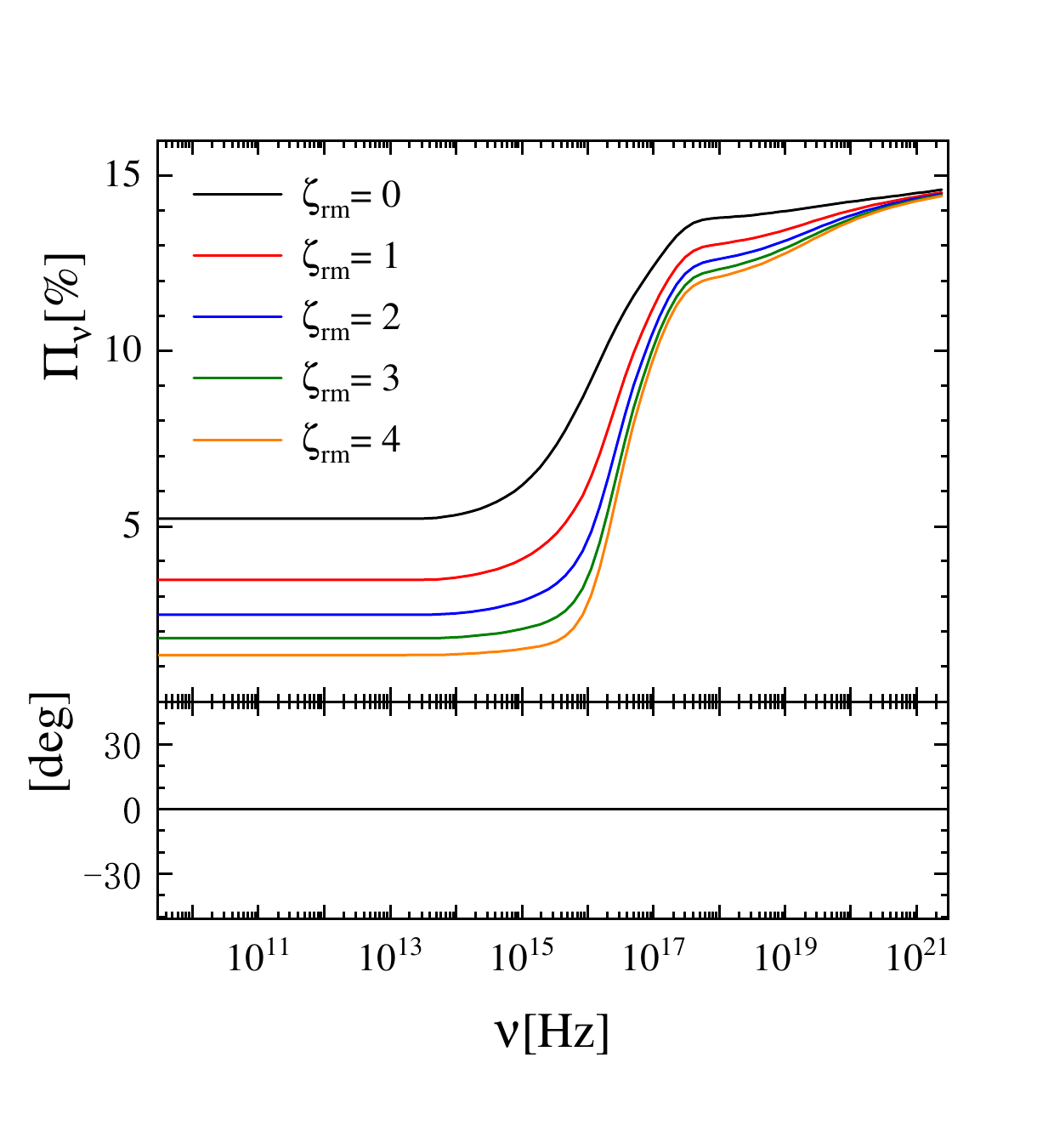}\\  
\caption{Same as Figure \ref{MyFigE} but with the sgMFs' decay index $\alpha_{\rm B}$=0.5.
The black, red, blue, green, and orange correspond to the cases of $\zeta_{\rm rm}$ = 0, 1, 2, 3, and 4, respectively.
Since the value of $\chi$ remains zero for all wave band and all studied cases in this figure,
we only show the case with $\zeta_{\rm rm}=0$ as an example.
}
\label{MyFigG}
\end{figure}


\begin{figure}
\centering
\includegraphics[width=0.8\textwidth]{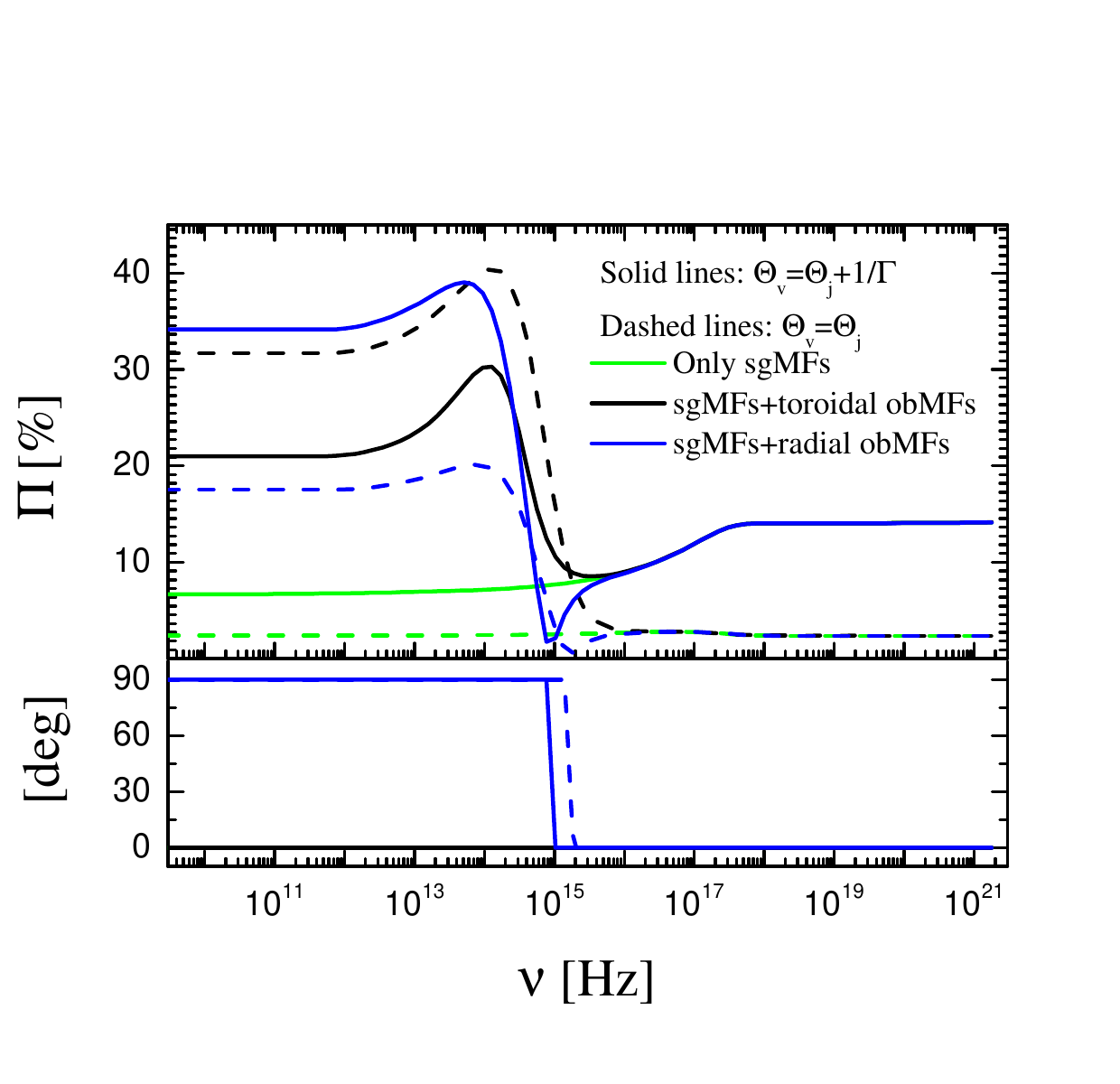}\\  
\caption{The $\nu-\Pi$ (upper panel) and $\nu-\chi$ (bottom panel) for with only sgMFs + obMFs,
where the solid and dashed lines are for $\Theta_{\rm v}=\Theta_{\rm j}+1/\Gamma$ and $\Theta_{\rm v}=\Theta_{\rm j}$, respectively.
The green, black, and blue lines are for the situation with only sgMFs,
sgMFs + toroidal obMFs, and sgMFs + radial obMFs, respectively.}
\label{MyFigD}
\end{figure}

\begin{figure}
\centering
 \hspace*{0.0truecm}
 \vspace*{0.0truecm}
\includegraphics[width=0.8\textwidth]{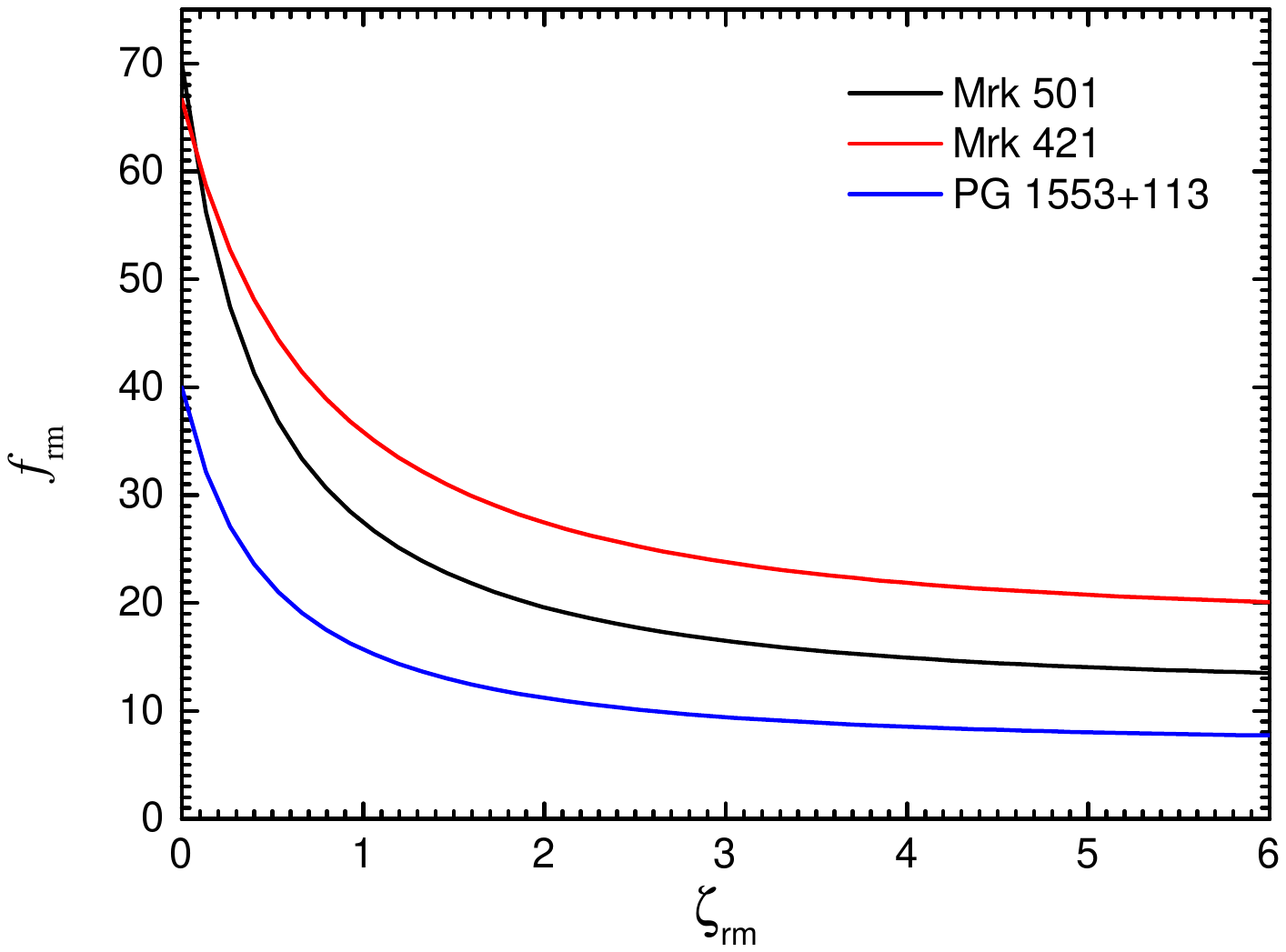}\\
\caption{The $\zeta_{\rm rm}$-dependent $f_{\rm rm}$ for discussed different blazars in Section~4.}
\label{MyFigH}
\end{figure}
\clearpage

\acknowledgments{
This work is supported by the National Natural Science Foundation of China
(grant Nos. 12273005 and 12133003), the Guangxi Science Foundation (grant Nos. 2018GXNSFFA281010),
and China Manned Spaced Project (CMS-CSST-2021-B11).}

\clearpage
\bibliographystyle{apj}
\bibliography{bibliography}
\end{document}